\documentclass[reprint,
superscriptaddress,
amsmath,amssymb,
aps,
prl,
]{revtex4-2}

\usepackage{graphicx}
\usepackage{dcolumn}
\usepackage{bm}
\usepackage{hyperref}
\usepackage{xcolor}
\usepackage{ulem}
\usepackage{xcolor}
\usepackage{soul}

\begin{document}

\preprint{APS/123-QED}
\title{Utilizing time-series measurements for entropy production estimation in partially observed systems}

\author{Uri Kapustin}
\affiliation{School of Electrical Engineering, Faculty of Engineering, Tel Aviv University, Tel Aviv 6997801, Israel}
\author{Aishani Ghosal}
\affiliation{Department of Biomedical Engineering, Faculty of Engineering, Tel Aviv University, Tel Aviv 6997801, Israel}
\author{Gili Bisker}%
\email{bisker@tauex.tau.ac.il}
\affiliation{Department of Biomedical Engineering, Faculty of Engineering, Tel Aviv University, Tel Aviv 6997801, Israel}
\affiliation{The Center for Physics and Chemistry of Living Systems, Tel Aviv University, Tel Aviv 6997801, Israel}
\affiliation{The Center for Nanoscience and Nanotechnology, Tel Aviv University, Tel Aviv 6997801, Israel}
\affiliation{The Center for Light Matter Interaction, Tel Aviv University, Tel Aviv 6997801, Israel}


\begin{abstract}
Estimating the dissipation, or the entropy production rate (EPR), can provide insights into the underlying mechanisms of nonequilibrium driven processes. Experimentally, however, only partial information can be accessed, and the ability to estimate the EPR varies depending on the available data. Here, we test different degrees of observed information stemming from coarse-grained time-series trajectory data, and apply several EPR estimators. Given increasing amount of information, we show a hierarchy of lower bounds on the total EPR. Further, we present a novel approach for utilizing waiting times in hidden states to provide a tighter lower bound on the total EPR.
\end{abstract}

\maketitle
    

The entropy production, or energy dissipation, is a fundamental physical quantity necessary to characterize the thermodynamics of nonequilibrium processes \cite{manikandan2022estimate,gnesotto2018broken}. In living systems, for example, the dissipation rate is closely related to the consumption rate of chemical fuel molecules, such as Adenosine triphosphate (ATP), by molecular motors \cite{molecularMotorExample}. 
The entropy production calculated along a single trajectory is a stochastic quantity that follows a set of mathematical relations, collectively known as the fluctuation theorems \cite{ciliberto2013heat,FT_Espo,FT_Seifert,evans2002fluctuation,sevick2008fluctuation,FT_Sagawa,FT_Espo_sum}, which have been validated experimentally \cite{zamponi2007possible,wang2002experimental,collin2005verification}.

Estimating the total EPR from experimental data for a driven system is not always trivial, and can be challenging owing to the limited resolution and the huge number of degrees of freedom \cite{gnesotto2018broken}. While calculating the EPR is straightforward given complete information about the nonequilibrium degrees of freedom, practically, only partial information is available. Such coarse-grained observation, where only some of the degrees of freedom are monitored or resolved, often cannot be treated with the Markovian approximation 
\cite{semiMarkov,semiMarkov2,Rahav_2007,Godec_KineticHysteresis}, and can only provide a lower bound on the total dissipation \cite{Esposito_CG_2012,Kawai_DissipationPhaseSpace,Fakhri_Jordan_2019,Bisker_2017}.

Partial information can stem from an observed sub-system, such that the rest of the system is hidden, or from coarse-graining some of the microstates into several meso-states \cite{Bisker_2017,CGexamples,CGexamples2,EPR_CG,Teza_ExactCG_PhysRevLett.125.110601}. 
The observed information may also be the transitions between states rather than the states themselves \cite{TransitionBased_Roldan,TransitionBased_Seifrit,Seifrit2022}.
Therefore, different levels of coarse-graining can be considered based on the partial information available about the system. 
 
There are several estimators for {partial} EPR that do not require any prior information about the system, such as the number of states or the underlying topology. The thermodynamic uncertainty relations (TUR), for example, provide a lower bound on the entropy production from the fluctuations of the transition fluxes or first passage times
\cite{shiraishi2021optimal,horowitz2020thermodynamic,falasco2020unifying,barato2015thermodynamic,TUR_ML_Sagawa,TUR_seifrit2,TURforSemiMarkov_Hasegawa,TURforSemiMarkov_Seifrit,TURderivation_Sagawa,TURderivation_Shiraishi,TURnature,GeneralizedTUR}.

Another approach for inferring a lower bound on the total EPR is based on an optimization problem, searching over systems with known topology and the same observed statistics, preserving the first and second-order mass transfer rates \cite{Dunkel_OPT}, or the waiting time statistics \cite{Dunkel_WTD}, or both \cite{Eden2022}.
A similar approach was also demonstrated for discrete-time Markov chains, searching over the possible underlying hidden Markov models given the number of hidden states \cite{AprioriEstimator}.
 
Building on the deep connection between the dissipation and the breaking of time-reversal symmetry \cite{parrondo2009arrow}, many estimators rely on the direct link between the EPR and the difficulty of distinguishing between forward and reverse processes, quantified by the relative entropy, or the Kullback–Leibler Divergence (KLD), between them  \cite{GiliNat19,Bisker_2017,KldCalcInMarkov,FR_semiCG_semiAnalytical,roldanFR,AishaniEPR,Ro_PhysRevLett.129.220601_Active_Matter}. 
Calculating the KLD between probability distributions of forward and reverse trajectories for stationary data series can be done using different approaches \cite{GiliNat19,FR_semiCG_semiAnalytical,roldanFR,Wang2005DivergenceEO,KldCalcInMarkov}. 
For example, the Plug-in method requires estimating the probabilities of sequences of data, discarding the information about transition times \cite{FR_semiCG_semiAnalytical}.

Applied to semi-Markov processes, the KLD breaks into two contributions, one of which captures irreversibility in the sequence of states, and the other captures irreversibility in waiting time distributions (WTD) \cite{GiliNat19}.  
For second-order semi-Markov processes, this KLD estimator
can detect and quantify entropy production even in the absence of observable currents \cite{GiliNat19,GiliCommentonGodec,GodecCommentonGili}.
Machine-Learning (ML) tools have also been used for entropy production estimation by exploiting the irreversibility of data series \cite{NEEP,NEEP2,NEEP3_Sagawa,EPR_traj}.
The core idea is to optimize an objective function whose extremum is the KLD between the forward and reverse trajectories of sequences of states. For example, the recurrent neural network estimator for entropy production (RNEEP) estimates the EPR from coarse-grained data of partially observed systems, using a recurrent neural network to solve the optimization problem \cite{NEEP}.

\begin{figure*}[t]
\includegraphics{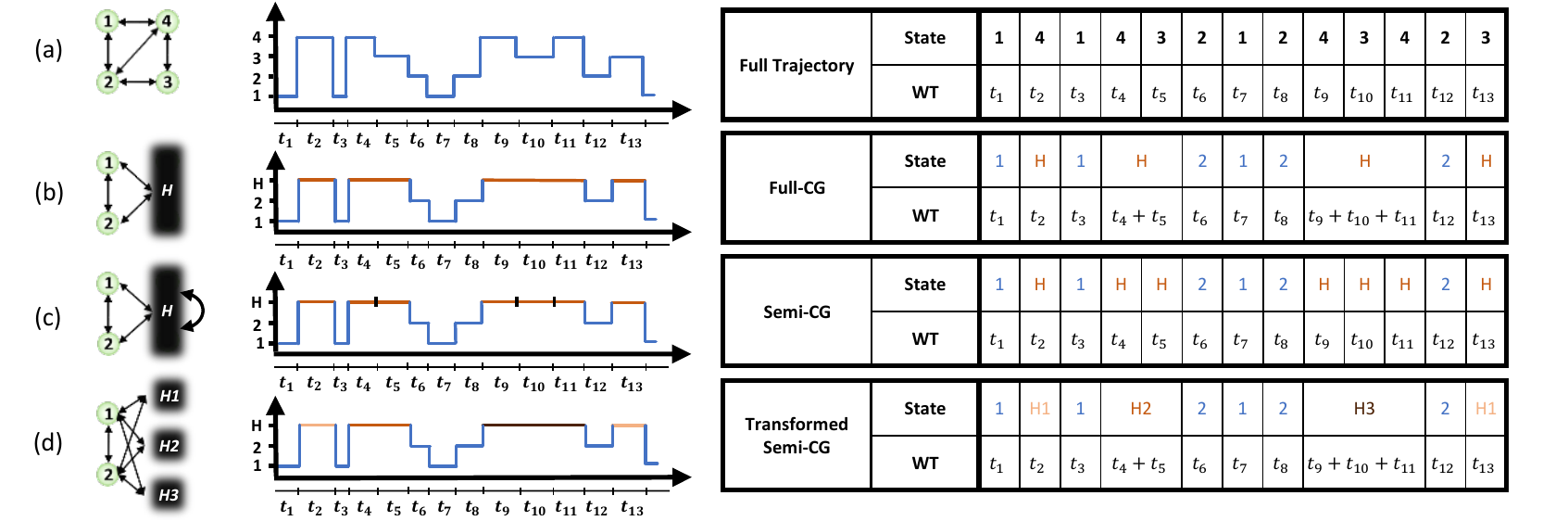}
\caption{\label{fig:frame}Illustration of the partial information frameworks for arbitrary 4-states system.
(a) A fully-observed 4-states system. The trajectory (blue line) is described by the sequence of microstates and the corresponding waiting times (WT). (b) Full coarse graining (\textit{full-CG}). States $3$ and $4$ cannot be resolved, and are lumped together to a single macrostate $H$ ({orange line})
(c) Semi-coarse graining (\textit{semi-CG}). States $3$ and $4$ cannot be resolved, but intra-transitions between the hidden microstates can be recorded,
where consecutive visits in the hidden microstates (orange line with black markers) are recorded as sequence of $H$, with the corresponding WT of each hidden microstate between intra-transition events. (d) Transformed Semi-coarse graining (\textit{Transformed semi-CG}).
Each $n$ consecutive visits to the hidden microstates in $H$ are recorded as $H_n$ (light orange, orange, and brown represent different sequence lengths) and the WT in $H_n$ are the sum of the WT in consecutive visits in the hidden microstates 3 and 4.}
\end{figure*}

\begin{figure}[t]
\includegraphics{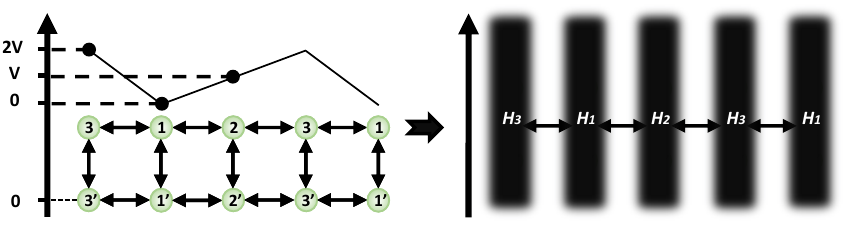}
\caption{\label{fig:models} Discrete flashing ratchet of $3$ states with periodic boundaries, and a potential that can be switched on ($i$) and off ($i'$) but is not accessible to the observer. State of the same position $i$, regardless of the potential, are coarse-grained into a single macrostate $H_1$, $H_2$, and $H_3$. In the \textit{semi-CG} framework, intra-transitions $i\leftrightarrow i'$ are recorded.}
\end{figure}

In this work, we focus on a continuous-time Markov chain (CTMC) model over a discrete set of states, in which a subset of the microstates are coarse-grained, or \textquotedblleft lumped\textquotedblright, into a single macrostate. We consider different levels of observed statistics
from different coarse-graining (CG) approaches, and infer the EPR from the observed data using the KLD estimator \cite{GiliNat19}, the Plug-in estimator \cite{roldanFR,FR_semiCG_semiAnalytical}, and the RNEEP estimator \cite{NEEP}, when applicable.  These estimators do not require prior knowledge of the systems, and only use the observed statistics to infer and quantify time-irreversibility. First, we use the sequence of observed microstates and coarse-grained macrostates, and the transitions between them.
Then, we include information about transitions between the hidden microstates within the coarse-grained macrostates (intra-transitions). Finally, we reformulate the trajectory data of observed states and transitions, and intra-transitions within macrostates, by labeling the coarse-grained macrostates according to the number of times they are visited before jumping 
into an observed state. We apply the CG approaches to two model systems, namely, a $4$-state {(Fig.~\ref{fig:frame})} system in which two of the states are coarse-grained into a single hidden state, and the discrete Flashing Ratchet {(Fig.~\ref{fig:models})} with time-varying potential, whose values cannot be observed.
We provide a unifying comparison between the estimators on the different CG schemes, and show how additional information is exploited for inferring tighter  lower bounds on the total EPR.


We begin by explaining the different coarse-graining approaches, taking the $4$-state model system as an example (Fig.~\ref{fig:frame}a). In the first CG approach, termed \textit{full-CG}, we lump together a subset of the microstates into a single observed state, giving rise to a second-order semi-Markov process (Fig.~\ref{fig:frame}b), since the waiting time in the hidden state depends on the state visited before \cite{GiliNat19}. In this example, states $1$ and $2$ are observed, whereas states $3$ and $4$ can not be distinguished and are recorded as a single state, $H$. Here, the waiting time in $H$ is the sum of the corresponding waiting times in the microstates $3$ and $4$ before jumping to one of the observed states.

In the second CG approach, termed \textit{semi-CG}, we assume an observer can record intra-transitions within the hidden states (Fig.~\ref{fig:frame}c). For example, a sequence of $1\rightarrow4\rightarrow3\rightarrow2$ is recorded as $1\rightarrow H\rightarrow H\rightarrow2$, with the corresponding waiting times, \textit{i.e.}, the time spent in the first visit to $H$ and the time spent in the second visit to $H$ are recorded separately. In this case of observed intra-transitions within a coarse-grained state, the initial and final microstates are not known, as both are lumped together to the same macrostate. 
Still, the added information can be utilized for improving the lower bound on the total EPR.

The Plug-in and the RNEEP estimators rely on
the sequence of states, and can be directly applied to the \textit{semi-CG} trajectory. However, in %
order to apply the KLD estimator to the \textit{semi-CG} data, we first need to reformulate the trajectory to a second-order semi-Markov process, and harness the information of the WTD. The transformation, depicted in Fig.~\ref{fig:frame}d, consists of two steps. First, we look for all the consecutive sequences of the hidden state $H$, and record their length. Second, all sequences with the same length are considered a new state, \textit{i.e.}, a sequence of $n$ appearances of $H$ is labeled $H_n$. The waiting time associated with $H_n$ is now the sum of the individual waiting times in the $n$ consecutive appearances of $H$. This new representation of the \textit{semi-CG} observed data gives rise to a second-order semi-Markov process, from which we infer a tighter lower bound on the total EPR. 

The Plug-in estimator, $\sigma_{\text{plug}}$, was proposed for approximating the KLD rate between the forward and reverse sequences of discrete stationary time series, by counting sequences of data and calculating their probabilities \cite{roldanFR,FR_semiCG_semiAnalytical}. 
The approximated {$m$th-order}
KLD between sequences of length $m$ is:
\begin{equation}
D^x_m =\displaystyle\sum_{x_1,x_2,...,x_n}p(x_{1\rightarrow m}){\ln}\left(\frac{p(x_{1\rightarrow m})}{p(x_{m\rightarrow 1})}\right)
\end{equation}
where ${p}(x_{1\rightarrow m})$ and ${p}(x_{m\rightarrow 1})$ are the probabilities of a forward sequence $x_{1\rightarrow m}=(x_1,...,x_m)$ and the backward one $x_{m\rightarrow 1}=(x_m,...,x_1)$. These probabilities can be estimated from the number of appearances of each sequence in a long trajectory.
Based on the approach in \cite{FR_semiCG_semiAnalytical}, the slope of $D^x_m$ as a function of m,
\begin{equation}
\hat{d}^x_m = D^x_m - D^x_{m-1}
\end{equation}
gives the entropy production per step in the limit of large $m$. 
However, for a non-Markov process that cannot be described by a semi-Markov process of any order, calculating $\hat{d}^x_m$ is challenging for large values of $m$. Therefore, the following ansatz \cite{pluginAnszats} has been proposed: 
\begin{equation}
\label{eqPLGN}
\hat{d}^x_m \simeq \hat{d}^x_\infty - c\frac{\ln(m)}{m^\gamma}
\end{equation}
where $\hat{d}^x_\infty$, $c$, and $\gamma$, are the fit parameters for $\hat{d}^x_m$ as a function of $m$.
Our Plug-in estimator for the entropy production rate \textit{per time}, is thus:
\begin{equation}
\label{eqPLGN_per_time}
\sigma_{\text{plug}} =\frac{1}{\tau} \hat{d}^x_\infty
\end{equation}
where $\tau$ is the mean waiting time in each step.
Note that this estimator can be directly used for both \textit{semi-CG} and \textit{full-CG} partial information framework without any modifications
to the trajectory data.

The KLD estimator, $\sigma_{\text {KLD}}$, derived by calculating the KLD between forward and reverse trajectories in semi-Markov processes, has two contributions \cite{GiliNat19}:
\begin{equation}
\label{eqKLD}
\sigma_{\text {KLD}} =  \sigma_{\text{aff}} + \sigma_{\text{WTD}}
\end{equation}
where the affinity, $\sigma_{\text{aff}}$, stems from observed currents, and the $\sigma_{\text{WTD}}$ stems from time-asymmetries in WTD.
In order to apply Eq. \ref{eqKLD} to second-order semi-Markov processes, the observed states are reformulated as doublets, $[ij]$, where the first index is the previous state, and the second index is the current state \cite{GiliNat19}. The affinity contribution is:
\begin{equation}
\label{eqKLDaff}
\sigma_{\text{aff}} = \frac{1}{\tau}\sum_{i,j,k}p_{(ijk)} \ln \left( \frac{p_{([ij]\rightarrow [jk])}}{p_{([kj]\rightarrow [ji])}} \right)
\end{equation}
where $p_{(ijk)}$ is the probability to observe the sequence of state $i\rightarrow j \rightarrow k$, or 
$p_{(ijk)}=p_{([ij]\rightarrow [jk])}R_{[ij]}$, with $p_{([ij]\rightarrow [jk])}$ being the probability to jump to state $k$ after jumping from $i$ to $j$, and $R_{[ij]}$ being the fraction of visits to $[i,j]$. The affinity, $\sigma_{\text{aff}}$, is governed by the relation between the forward and reverse transition probabilities. 

The WTD contribution stems from the Kullback-Leibler divergence between WTD associated with forward ($i\rightarrow j\rightarrow k$) and backward ($k\rightarrow j\rightarrow i$) transitions:
\begin{equation}
\label{eqKLDwtd}
\sigma_{\text{WTD}} = \frac{1}{\tau}\displaystyle\sum_{i,j,k}p_{(ijk)} D\left[\psi(t|[ij]\rightarrow [jk] )||\psi(t|
[kj]\rightarrow[ji])\right]
\end{equation}
where $\psi (t|[ij]\rightarrow [jk] )$ is the WTD in state $j$ 
given that the previous state was $i$ and the following is $k$, $\tau$ is the average waiting time per state, and $D[u(x)||v(x)]$ is the Kullback-Leibler Divergence between two probability distributions, $u(x)$ and $v(x)$, defined as $D[u(x)||v(x)]= \sum_{x} u(x)\ln\left(u(x)/v(x)\right)$.
See Supplemental Material (SM) \cite{SI} for details regarding WTD estimation.
Note that for a fully observed system following Markovian dynamics, $\sigma_{\text{WTD}}$ vanishes, and one cannot infer non-zero EPR without observable currents.

\begin{figure*}[t]
\includegraphics{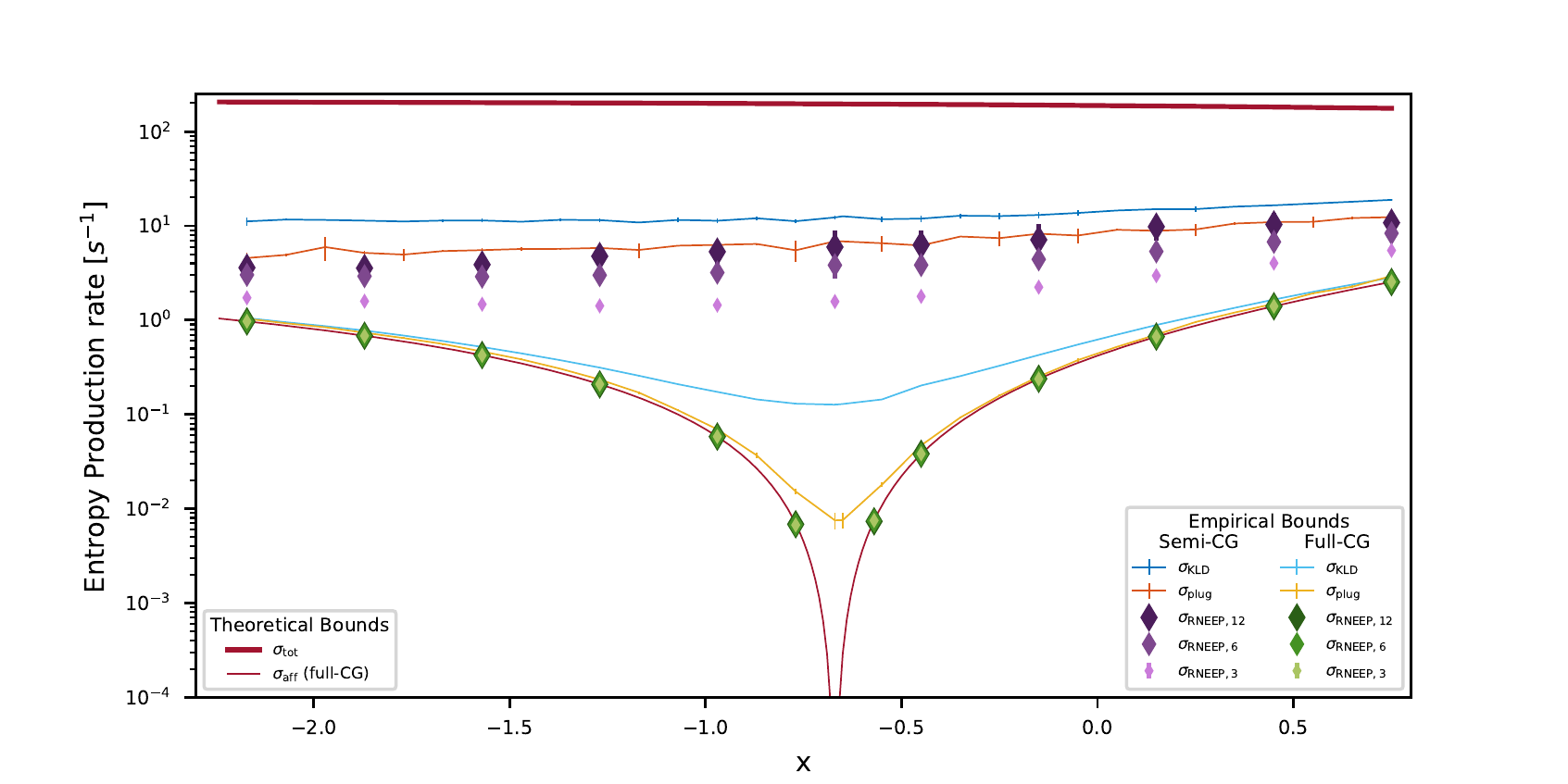}
\caption{\label{fig:res4s}
Entropy production rates for the $4$-state system. Total EPR, $\sigma_{\text{tot}}$ (dark red), KLD estimator, $\sigma_{\text{KLD}}$, for the \textit{semi-CG} (dark blue) and \textit{full-CG} (light blue) data, Plug-in estimator, $\sigma_{\text{plug}}$, for the \textit{semi-CG} (dark orange) and \textit{full-CG} (light orange), RNEEP estimator, $\sigma_{\text{RNEEP},m}$, for the \textit{semi-CG} (light to dark purple for increasing sequence length $m$) and \textit{full-CG} (light to dark green for increasing sequence length $m$) data, and the affinity contribution, $\sigma_{\text{aff}}$, for the \textit{full-CG} data (red).
Rates:  $\bar\omega_{12}=2s^{-1}$, 
$\bar\omega_{21}=3s^{-1}$, 
$\omega_{13}=0s^{-1}$, $\omega_{14}=1s^{-1}$, $\omega_{23}=2s^{-1}$, $\omega_{24}=35s^{-1}$, $\omega_{31}=0s^{-1}$, $\omega_{32}=50s^{-1}$, $\omega_{34}=0.7s^{-1}$, $\omega_{41}=8s^{-1}$, $\omega_{42}=0.2s^{-1}$, $\omega_{43}=75s^{-1}$. 
}
\end{figure*}

The RNEEP estimator{,} $\sigma_{\text{RNEEP}}${,} is formulated as an optimization problem \cite{NEEP}, with a specific objective function to be {minimized} using stochastic gradient {descent.}
The input of the problem is the set of all sequences of length $m$ from a single long trajectory,
and the solution is the coarse-grained entropy production rate per step along the input trajectory.
Similar to the Plug-in estimator, the RNEEP uses the discrete sequence of states and does not exploit the WTD data, so estimating the full probability distributions of the waiting times is not required.
Intuitively, this estimator should yield similar results to the Plug-in estimator, Eq.{~\ref{eqPLGN_per_time}}, and to $\sigma_{\text{aff}}$, Eq.~\ref{eqKLDaff}, as it uses the same information (see SM for further discussion \cite{SI}).
The RNEEP can be directly applied to both \textit{full-CG} and \textit{semi-CG} frameworks, and it can be implemented by different machine learning models, such as recurrent or convolutional neural networks \cite{NEEP2,NEEP}.
Following the approach of \cite{NEEP}, we use a recurrent neural network, whose input is a sequence of some length $m$, $x^m_t = (x_t,x_{t+1},...,x_{t+m-1})$, and its output is $h_\theta (x^m_t)$, where $\theta$ represents the learnable weights of the network. The output of the RNEEP is \cite{NEEP}:
\begin{equation}
\Delta S_\theta (x^m_t) \equiv h_\theta (x^m_t) - h_\theta (\tilde{x}^m_t)
\end{equation}
where $\tilde{x}^m_t$ is the time-reversed sequence of $x^m_t$.
The RNEEP estimator is the solution of the optimization problem of {minimizing} the following objective function over $\Delta S_\theta (x^m_t)$ for all possible sequences of length $m$:
\begin{equation}
    \sigma_{\text{RNEEP},m} = \frac{1}{\tau} \displaystyle\min_{\Delta S_\theta}\mathrm{E}_t\mathrm{E}_{(x^m_t)}\small{\left[   \Delta S_\theta (x^m_t) - e^{-\Delta S_\theta (x^m_t)}   \right]}
\end{equation}
where $\mathrm{E}_t$ in the expectation over $t$, and $\mathrm{E}_{(x^m_t)}$ is the expectation over the observed sequences $x^m_t$. See SM \cite{SI} for a detailed explanation of the implementation of the estimators and the different numerical considerations.

We have evaluated the performance of the three EPR estimators on two coarse-grained systems, the $4$-state system with {1 coarse-grained} state, and the discrete flashing ratchet with the unobserved external potential. For each system, the two CG approaches, namely, the \textit{full-CG} and \textit{semi-CG}, were applied on trajectories of approximately $N=10^7$ states, simulated using the Gillespie algorithm \cite{Gillespie}.
The code was implemented in PyTorch is available in \cite{code}.

The Plug-in estimator was fitted by gathering statistics of sequences of various lengths 
according to Eq. \ref{eqPLGN}, as done in \cite{FR_semiCG_semiAnalytical}. The RNEEP was calculated for different input sizes, where we have used the implementation of D.-K. Kim, \textit{et. al.} \cite{NEEP} with some adjustments for our hardware, see \cite{SI}. The KLD estimator was applied without modifications on the \textit{full-CG} data \cite{GiliNat19}, whereas the trajectory reformulation was used only for calculating $\sigma_{\text{KLD}}$ for the \textit{semi-CG} statistics.


The results for the 4-states system (Fig.~\ref{fig:frame}(a)) under the two CG schemes, \textit{semi-CG} and \textit{full-CG}, and the three estimators, RNEEP, Plug-in, and KLD, are presented in Fig.~\ref{fig:res4s}. 
The system has two observed states, $1$, and $2$, where the states $3$ and $4$ are coarse-grained into a single state $H$. The rates between the two observed microstates are tuned according to  $\omega_{12}=\bar\omega_{12}e^{x}$ and $\omega_{21}=\bar\omega_{21}e^{-x}$ to mimic an external forcing, where the range of $x$ was chosen to include the stalling force in which there is no observable current over the $1-2$ link \cite{Bisker_2017}.

\begin{figure}[t]
\includegraphics{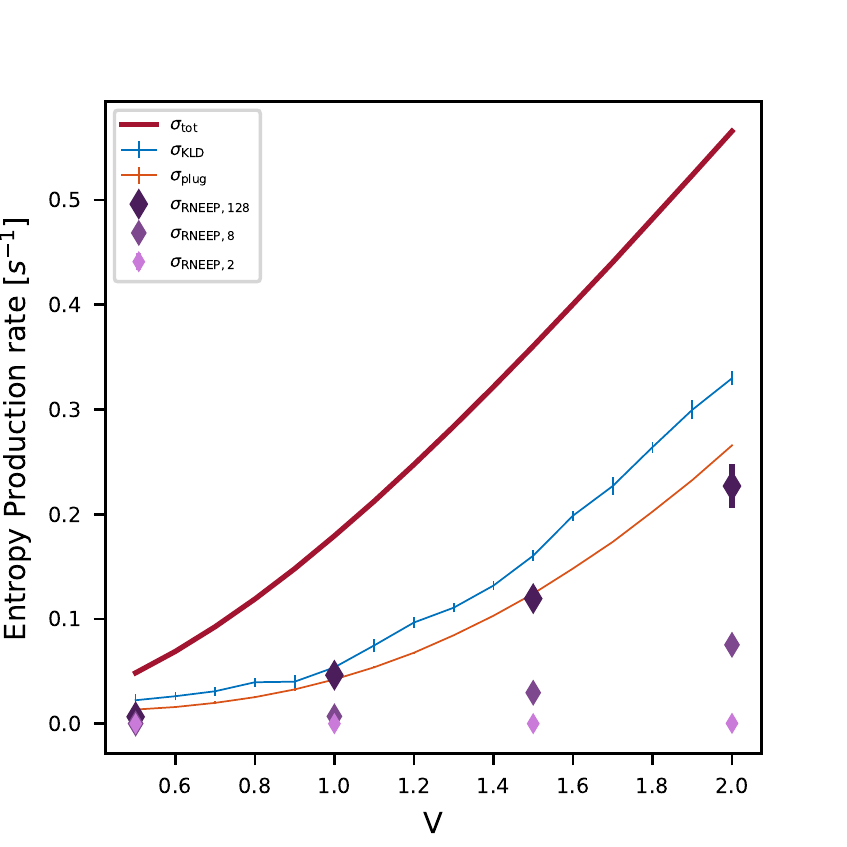}
\caption{\label{fig:resFR} {Entropy} production rates for the flashing ratchet. Total EPR, $\sigma_{\text{tot}}$ (dark red) of the full trajectory,
KLD estimator, $\sigma_{\text{KLD}}$ (dark blue), Plug-in estimator, $\sigma_{\text{plug}}$ (dark orange), and RNEEP estimator, $\sigma_{\text{RNEEP},m}$ (light to dark purple for increasing sequence length $m$) for \textit{semi-CG} data.
The transition rates for $\forall i\neq j$ are: $\omega_{ii'}=\omega_{i'i}=\omega_{i'j'}=1$, $\omega_{ij}=e^{(V_j-V_i)/2}$, where states $i$ ($i'$) are with turned-on (off) potential.
}
\end{figure}

As expected, The bounds on the total EPR obtained from estimators applied to the \textit{semi-CG} statistics are better compared with the same estimators applied to the \textit{full-CG} trajectories. In the \textit{full-CG} case, $\sigma_{\text{RNEEP},m}$ {is similar to} the affinity, $\sigma_{\text{aff}}$, as both estimators use the same data.
The Plug-in estimator, $\sigma_{\text{plug}}$, which also uses the same data of the \textit{full-CG} trajectory, provides similar results to $\sigma_{\text{aff}}$ away from the stalling force. However, close to the stalling force, where $\sigma_{\text{aff}}$ vanishes, $\sigma_{\text{plug}}$ provides non-zero values that stem from the inherent bias of the method which assigns positive values to all the probabilities \cite{roldanFR} (See {\cite{SI}}). 
The KLD estimator, $\sigma_{\text{KLD}}$, provides the tightest lower bound for the \textit{full-CG} data, as it is the only one that utilizes information of the irreversibility in WTD.

In the \textit{semi-CG} case, the RNEEP estimator, $\sigma_{\text{RNEEP},m}$, provides a tighter bound for increasing sequence lengths $m$, 
and it converges to $\sigma_{\text{plug}}$ as in the \textit{full-CG} case \cite{SI}.
The KLD estimator $\sigma_{\text{KLD}}$, provides the tightest lower bound on the total EPR compared to the other estimators tested, given the reformulation of the trajectory data according to the \textit{Transformed semi-CG} scheme (Fig.~\ref{fig:frame}(d)). The difference in $\sigma_{\text{KLD}}$ between the \textit{Transformed semi-CG} and the \textit{full-CG} schemes reflects the additional information regarding irreversibility encoded in the intra-transitions between microstates in the hidden macrostate.
Moreover, the irreversibility encoded in these intra-transitions in the \textit{semi-CG} data is also reflected in the values of the $3$ estimators that do not vary significantly near the stall force, in contrast to the estimators applied to the \textit{full-CG} trajectories that strongly depend on the deviation from stalling conditions.

The results for the discrete flashing ratchet system \cite{roldanFR,FR_semiCG_semiAnalytical} under the \textit{semi-CG} scheme and the three estimators, RNEEP, Plug-in, and KLD, are presented in Fig.~\ref{fig:resFR}. The system is of a Brownian particle moving along a periodic one-dimensional line, under the influence of a linear potential $V$ that can be switched on and off at a constant rate. The particle is described by its position in the ``on'', $i$, or ``off'', $i'$, states. 
Under CG, the information on the potential is not accessible, and both the on and off states are lumped into a single macrostate $H_i$.
Here, we apply the $3$ estimators to the non-Markovian \textit{semi-CG} data, which includes the information of the intra-transition within the macrostates.
Note that in \cite{NEEP}, the RNEEP was compared to a semianalytical calculation of the KLD between trajectory distributions. However, it was shown that the Plug-in estimator yielded similar results to the semianalytical values for the \textit{semi-CG} observed statistics \cite{FR_semiCG_semiAnalytical}.
Similar to the 4-state system results, the RNEEP estimator, $\sigma_{\text{RNEEP},m}$, provides a tighter bound for increasing sequence lengths $m$.
Moreover, the application of the KLD estimator to the \textit{Transformed semi-CG} date yields the tightest lower bound on the total EPR, compared to the Plug-in \cite{roldanFR,FR_semiCG_semiAnalytical} and the RNEEP \cite{NEEP}.


In summary, we have compared time-irreversibility-based EPR estimators for different coarse-graining schemes, focusing on KLD-based estimators, with (KLD) and without (Plug-in and RNEEP) WTD statistics, 
using two coarse-graining approaches.
We have confirmed that the semi-coarse-graining framework, which includes intra-transitions data, yields tighter EPR bounds compared to the full-coarse-graining framework, as it exploits more information on time-irreversibility.
In addition, we have proposed a novel approach for reformulating \textit{semi-CG} trajectories, previously used for the Plug-in \cite{roldanFR,FR_semiCG_semiAnalytical} and RNEEP estimators \cite{NEEP}, for applying the KLD estimator \cite{GiliNat19}.
Using the {\textit{Transformed semi-CG}}
approach and the KLD estimator, we could distill time-irreversibility encoded in the intra-transitions within hidden microstates to achieve the tightest lower bound on the total EPR among the estimators we tested.  Moreover, comparing the EPR bounds obtained from the \textit{full-CG} and the \textit{semi-CG} statistics provides a direct quantification of the time-irreversibility in the intra-transitions captured by each estimator.
The proposed transformation and EPR estimators can be applied to other discrete-state continuous-time systems, to provide a lower bound on the total EPR, when only partial information is available.
\\
\\
G. Bisker acknowledges the Zuckerman STEM Leadership Program, and the Tel Aviv University Center for AI and Data Science (TAD). This work was supported
by the ERC NanoNonEq 101039127, the Air Force Office of Scientific Research (AFOSR) under award number FA9550-20-1-0426, and by the Army Research Office
(ARO) under Grant Number W911NF-21-1-0101. The
views and conclusions contained in this document are
those of the authors and should not be interpreted as representing the official policies, either expressed or implied,
of the Army Research Office or the U.S. Government.

\bibliography{papers}

\providecommand{\noopsort}[1]{}\providecommand{\singleletter}[1]{#1}%
\begin{thebibliography}{62}%
\makeatletter
\providecommand \@ifxundefined [1]{%
 \@ifx{#1\undefined}
}%
\providecommand \@ifnum [1]{%
 \ifnum #1\expandafter \@firstoftwo
 \else \expandafter \@secondoftwo
 \fi
}%
\providecommand \@ifx [1]{%
 \ifx #1\expandafter \@firstoftwo
 \else \expandafter \@secondoftwo
 \fi
}%
\providecommand \natexlab [1]{#1}%
\providecommand \enquote  [1]{``#1''}%
\providecommand \bibnamefont  [1]{#1}%
\providecommand \bibfnamefont [1]{#1}%
\providecommand \citenamefont [1]{#1}%
\providecommand \href@noop [0]{\@secondoftwo}%
\providecommand \href [0]{\begingroup \@sanitize@url \@href}%
\providecommand \@href[1]{\@@startlink{#1}\@@href}%
\providecommand \@@href[1]{\endgroup#1\@@endlink}%
\providecommand \@sanitize@url [0]{\catcode `\\12\catcode `\$12\catcode
  `\&12\catcode `\#12\catcode `\^12\catcode `\_12\catcode `\%12\relax}%
\providecommand \@@startlink[1]{}%
\providecommand \@@endlink[0]{}%
\providecommand \url  [0]{\begingroup\@sanitize@url \@url }%
\providecommand \@url [1]{\endgroup\@href {#1}{\urlprefix }}%
\providecommand \urlprefix  [0]{URL }%
\providecommand \Eprint [0]{\href }%
\providecommand \doibase [0]{https://doi.org/}%
\providecommand \selectlanguage [0]{\@gobble}%
\providecommand \bibinfo  [0]{\@secondoftwo}%
\providecommand \bibfield  [0]{\@secondoftwo}%
\providecommand \translation [1]{[#1]}%
\providecommand \BibitemOpen [0]{}%
\providecommand \bibitemStop [0]{}%
\providecommand \bibitemNoStop [0]{.\EOS\space}%
\providecommand \EOS [0]{\spacefactor3000\relax}%
\providecommand \BibitemShut  [1]{\csname bibitem#1\endcsname}%
\let\auto@bib@innerbib\@empty
\bibitem [{\citenamefont {Manikandan}\ \emph {et~al.}(2022)\citenamefont
  {Manikandan}, \citenamefont {Ghosh}, \citenamefont {Mandal}, \citenamefont
  {Biswas}, \citenamefont {Sinha},\ and\ \citenamefont
  {Mitra}}]{manikandan2022estimate}%
  \BibitemOpen
  \bibfield  {author} {\bibinfo {author} {\bibfnamefont {S.~K.}\ \bibnamefont
  {Manikandan}}, \bibinfo {author} {\bibfnamefont {T.}~\bibnamefont {Ghosh}},
  \bibinfo {author} {\bibfnamefont {T.}~\bibnamefont {Mandal}}, \bibinfo
  {author} {\bibfnamefont {A.}~\bibnamefont {Biswas}}, \bibinfo {author}
  {\bibfnamefont {B.}~\bibnamefont {Sinha}},\ and\ \bibinfo {author}
  {\bibfnamefont {D.}~\bibnamefont {Mitra}},\ }\bibfield  {title} {\bibinfo
  {title} {Estimate of entropy generation rate can spatiotemporally resolve the
  active nature of cell flickering},\ }\Eprint
  {https://arxiv.org/abs/2205.12849} {arXiv:2205.12849}  (\bibinfo {year}
  {2022})\BibitemShut {NoStop}%
\bibitem [{\citenamefont {Gnesotto}\ \emph {et~al.}(2018)\citenamefont
  {Gnesotto}, \citenamefont {Mura}, \citenamefont {Gladrow},\ and\
  \citenamefont {Broedersz}}]{gnesotto2018broken}%
  \BibitemOpen
  \bibfield  {author} {\bibinfo {author} {\bibfnamefont {F.~S.}\ \bibnamefont
  {Gnesotto}}, \bibinfo {author} {\bibfnamefont {F.}~\bibnamefont {Mura}},
  \bibinfo {author} {\bibfnamefont {J.}~\bibnamefont {Gladrow}},\ and\ \bibinfo
  {author} {\bibfnamefont {C.~P.}\ \bibnamefont {Broedersz}},\ }\bibfield
  {title} {\bibinfo {title} {Broken detailed balance and non-equilibrium
  dynamics in living systems: a review},\ }\href
  {https://doi.org/10.1088/1361-6633/aab3ed} {\bibfield  {journal} {\bibinfo
  {journal} {Reports on Progress in Physics}\ }\textbf {\bibinfo {volume}
  {81}},\ \bibinfo {pages} {066601} (\bibinfo {year} {2018})}\BibitemShut
  {NoStop}%
\bibitem [{\citenamefont {Parrondo}\ and\ \citenamefont
  {Cisneros}(2002)}]{molecularMotorExample}%
  \BibitemOpen
  \bibfield  {author} {\bibinfo {author} {\bibfnamefont {J.~M.}\ \bibnamefont
  {Parrondo}}\ and\ \bibinfo {author} {\bibfnamefont {B.}~\bibnamefont
  {Cisneros}},\ }\bibfield  {title} {\bibinfo {title} {Energetics of brownian
  motors: A review},\ }\href {https://doi.org/10.1007/s003390201332} {\bibfield
   {journal} {\bibinfo  {journal} {Applied Physics A}\ }\textbf {\bibinfo
  {volume} {75}},\ \bibinfo {pages} {179} (\bibinfo {year} {2002})}\BibitemShut
  {NoStop}%
\bibitem [{\citenamefont {Ciliberto}\ \emph {et~al.}(2013)\citenamefont
  {Ciliberto}, \citenamefont {Imparato}, \citenamefont {Naert},\ and\
  \citenamefont {Tanase}}]{ciliberto2013heat}%
  \BibitemOpen
  \bibfield  {author} {\bibinfo {author} {\bibfnamefont {S.}~\bibnamefont
  {Ciliberto}}, \bibinfo {author} {\bibfnamefont {A.}~\bibnamefont {Imparato}},
  \bibinfo {author} {\bibfnamefont {A.}~\bibnamefont {Naert}},\ and\ \bibinfo
  {author} {\bibfnamefont {M.}~\bibnamefont {Tanase}},\ }\bibfield  {title}
  {\bibinfo {title} {Heat flux and entropy produced by thermal fluctuations},\
  }\href {https://doi.org/10.1103/PhysRevLett.110.180601} {\bibfield  {journal}
  {\bibinfo  {journal} {Phys. Rev. Lett.}\ }\textbf {\bibinfo {volume} {110}},\
  \bibinfo {pages} {180601} (\bibinfo {year} {2013})}\BibitemShut {NoStop}%
\bibitem [{\citenamefont {Esposito}\ and\ \citenamefont {Van~den
  Broeck}(2010)}]{FT_Espo}%
  \BibitemOpen
  \bibfield  {author} {\bibinfo {author} {\bibfnamefont {M.}~\bibnamefont
  {Esposito}}\ and\ \bibinfo {author} {\bibfnamefont {C.}~\bibnamefont {Van~den
  Broeck}},\ }\bibfield  {title} {\bibinfo {title} {Three detailed fluctuation
  theorems},\ }\href {https://doi.org/10.1103/PhysRevLett.104.090601}
  {\bibfield  {journal} {\bibinfo  {journal} {Phys. Rev. Lett.}\ }\textbf
  {\bibinfo {volume} {104}},\ \bibinfo {pages} {090601} (\bibinfo {year}
  {2010})}\BibitemShut {NoStop}%
\bibitem [{\citenamefont {Seifert}(2012)}]{FT_Seifert}%
  \BibitemOpen
  \bibfield  {author} {\bibinfo {author} {\bibfnamefont {U.}~\bibnamefont
  {Seifert}},\ }\bibfield  {title} {\bibinfo {title} {Stochastic
  thermodynamics, fluctuation theorems and molecular machines},\ }\href
  {https://doi.org/10.1088/0034-4885/75/12/126001} {\bibfield  {journal}
  {\bibinfo  {journal} {Reports on Progress in Physics}\ }\textbf {\bibinfo
  {volume} {75}},\ \bibinfo {pages} {126001} (\bibinfo {year}
  {2012})}\BibitemShut {NoStop}%
\bibitem [{\citenamefont {Evans}\ and\ \citenamefont
  {Searles}(2002)}]{evans2002fluctuation}%
  \BibitemOpen
  \bibfield  {author} {\bibinfo {author} {\bibfnamefont {D.~J.}\ \bibnamefont
  {Evans}}\ and\ \bibinfo {author} {\bibfnamefont {D.~J.}\ \bibnamefont
  {Searles}},\ }\bibfield  {title} {\bibinfo {title} {The fluctuation
  theorem},\ }\href {https://doi.org/10.1080/00018730210155133} {\bibfield
  {journal} {\bibinfo  {journal} {Advances in Physics}\ }\textbf {\bibinfo
  {volume} {51}},\ \bibinfo {pages} {1529} (\bibinfo {year}
  {2002})}\BibitemShut {NoStop}%
\bibitem [{\citenamefont {Sevick}\ \emph {et~al.}(2008)\citenamefont {Sevick},
  \citenamefont {Prabhakar}, \citenamefont {Williams},\ and\ \citenamefont
  {Searles}}]{sevick2008fluctuation}%
  \BibitemOpen
  \bibfield  {author} {\bibinfo {author} {\bibfnamefont {E.}~\bibnamefont
  {Sevick}}, \bibinfo {author} {\bibfnamefont {R.}~\bibnamefont {Prabhakar}},
  \bibinfo {author} {\bibfnamefont {S.~R.}\ \bibnamefont {Williams}},\ and\
  \bibinfo {author} {\bibfnamefont {D.~J.}\ \bibnamefont {Searles}},\
  }\bibfield  {title} {\bibinfo {title} {Fluctuation theorems},\ }\href
  {https://doi.org/10.1146/annurev.physchem.58.032806.104555} {\bibfield
  {journal} {\bibinfo  {journal} {Annual Review of Physical Chemistry}\
  }\textbf {\bibinfo {volume} {59}},\ \bibinfo {pages} {603} (\bibinfo {year}
  {2008})},\ \bibinfo {note} {pMID: 18393680}\BibitemShut {NoStop}%
\bibitem [{\citenamefont {Shiraishi}\ and\ \citenamefont
  {Sagawa}(2015)}]{FT_Sagawa}%
  \BibitemOpen
  \bibfield  {author} {\bibinfo {author} {\bibfnamefont {N.}~\bibnamefont
  {Shiraishi}}\ and\ \bibinfo {author} {\bibfnamefont {T.}~\bibnamefont
  {Sagawa}},\ }\bibfield  {title} {\bibinfo {title} {Fluctuation theorem for
  partially masked nonequilibrium dynamics},\ }\href
  {https://doi.org/10.1103/PhysRevE.91.012130} {\bibfield  {journal} {\bibinfo
  {journal} {Phys. Rev. E}\ }\textbf {\bibinfo {volume} {91}},\ \bibinfo
  {pages} {012130} (\bibinfo {year} {2015})}\BibitemShut {NoStop}%
\bibitem [{\citenamefont {Polettini}\ and\ \citenamefont
  {Esposito}(2019)}]{FT_Espo_sum}%
  \BibitemOpen
  \bibfield  {author} {\bibinfo {author} {\bibfnamefont {M.}~\bibnamefont
  {Polettini}}\ and\ \bibinfo {author} {\bibfnamefont {M.}~\bibnamefont
  {Esposito}},\ }\bibfield  {title} {\bibinfo {title} {Effective fluctuation
  and response theory},\ }\href
  {https://link.springer.com/article/10.1007/s10955-019-02291-7} {\bibfield
  {journal} {\bibinfo  {journal} {Journal of Statistical Physics}\ }\textbf
  {\bibinfo {volume} {176}},\ \bibinfo {pages} {94} (\bibinfo {year}
  {2019})}\BibitemShut {NoStop}%
\bibitem [{\citenamefont {Zamponi}(2007)}]{zamponi2007possible}%
  \BibitemOpen
  \bibfield  {author} {\bibinfo {author} {\bibfnamefont {F.}~\bibnamefont
  {Zamponi}},\ }\bibfield  {title} {\bibinfo {title} {Is it possible to
  experimentally verify the fluctuation relation? a review of theoretical
  motivations and numerical evidence},\ }\href
  {https://doi.org/10.1088/1742-5468/2007/02/P02008} {\bibfield  {journal}
  {\bibinfo  {journal} {Journal of Statistical Mechanics: Theory and
  Experiment}\ }\textbf {\bibinfo {volume} {2007}},\ \bibinfo {pages} {P02008}
  (\bibinfo {year} {2007})}\BibitemShut {NoStop}%
\bibitem [{\citenamefont {Wang}\ \emph {et~al.}(2002)\citenamefont {Wang},
  \citenamefont {Sevick}, \citenamefont {Mittag}, \citenamefont {Searles},\
  and\ \citenamefont {Evans}}]{wang2002experimental}%
  \BibitemOpen
  \bibfield  {author} {\bibinfo {author} {\bibfnamefont {G.~M.}\ \bibnamefont
  {Wang}}, \bibinfo {author} {\bibfnamefont {E.~M.}\ \bibnamefont {Sevick}},
  \bibinfo {author} {\bibfnamefont {E.}~\bibnamefont {Mittag}}, \bibinfo
  {author} {\bibfnamefont {D.~J.}\ \bibnamefont {Searles}},\ and\ \bibinfo
  {author} {\bibfnamefont {D.~J.}\ \bibnamefont {Evans}},\ }\bibfield  {title}
  {\bibinfo {title} {Experimental demonstration of violations of the second law
  of thermodynamics for small systems and short time scales},\ }\href
  {https://doi.org/10.1103/PhysRevLett.89.050601} {\bibfield  {journal}
  {\bibinfo  {journal} {Phys. Rev. Lett.}\ }\textbf {\bibinfo {volume} {89}},\
  \bibinfo {pages} {050601} (\bibinfo {year} {2002})}\BibitemShut {NoStop}%
\bibitem [{\citenamefont {Collin}\ \emph {et~al.}(2005)\citenamefont {Collin},
  \citenamefont {Ritort}, \citenamefont {Jarzynski}, \citenamefont {Smith},
  \citenamefont {Tinôco},\ and\ \citenamefont
  {Bustamante}}]{collin2005verification}%
  \BibitemOpen
  \bibfield  {author} {\bibinfo {author} {\bibfnamefont {D.}~\bibnamefont
  {Collin}}, \bibinfo {author} {\bibfnamefont {F.}~\bibnamefont {Ritort}},
  \bibinfo {author} {\bibfnamefont {C.}~\bibnamefont {Jarzynski}}, \bibinfo
  {author} {\bibfnamefont {S.}~\bibnamefont {Smith}}, \bibinfo {author}
  {\bibfnamefont {I.}~\bibnamefont {Tinôco}},\ and\ \bibinfo {author}
  {\bibfnamefont {C.}~\bibnamefont {Bustamante}},\ }\bibfield  {title}
  {\bibinfo {title} {Verification of the crooks fluctuation theorem and
  recovery of rna folding free energies},\ }\href
  {https://doi.org/10.1038/nature04061} {\bibfield  {journal} {\bibinfo
  {journal} {Nature}\ }\textbf {\bibinfo {volume} {437}},\ \bibinfo {pages}
  {231} (\bibinfo {year} {2005})}\BibitemShut {NoStop}%
\bibitem [{\citenamefont {Wang}\ and\ \citenamefont {Qian}(2007)}]{semiMarkov}%
  \BibitemOpen
  \bibfield  {author} {\bibinfo {author} {\bibfnamefont {H.}~\bibnamefont
  {Wang}}\ and\ \bibinfo {author} {\bibfnamefont {H.}~\bibnamefont {Qian}},\
  }\bibfield  {title} {\bibinfo {title} {On detailed balance and reversibility
  of semi-markov processes and single-molecule enzyme kinetics},\ }\href
  {https://doi.org/10.1063/1.2432065} {\bibfield  {journal} {\bibinfo
  {journal} {Journal of Mathematical Physics}\ }\textbf {\bibinfo {volume}
  {48}},\ \bibinfo {pages} {013303} (\bibinfo {year} {2007})}\BibitemShut
  {NoStop}%
\bibitem [{\citenamefont {Maes}\ \emph {et~al.}(2009)\citenamefont {Maes},
  \citenamefont {Neto{\v{c}}n{\'{y}}},\ and\ \citenamefont
  {Wynants}}]{semiMarkov2}%
  \BibitemOpen
  \bibfield  {author} {\bibinfo {author} {\bibfnamefont {C.}~\bibnamefont
  {Maes}}, \bibinfo {author} {\bibfnamefont {K.}~\bibnamefont
  {Neto{\v{c}}n{\'{y}}}},\ and\ \bibinfo {author} {\bibfnamefont
  {B.}~\bibnamefont {Wynants}},\ }\bibfield  {title} {\bibinfo {title}
  {Dynamical fluctuations for semi-markov processes},\ }\href
  {https://doi.org/10.1088/1751-8113/42/36/365002} {\bibfield  {journal}
  {\bibinfo  {journal} {Journal of Physics A: Mathematical and Theoretical}\
  }\textbf {\bibinfo {volume} {42}},\ \bibinfo {pages} {365002} (\bibinfo
  {year} {2009})}\BibitemShut {NoStop}%
\bibitem [{\citenamefont {Rahav}\ and\ \citenamefont
  {Jarzynski}(2007)}]{Rahav_2007}%
  \BibitemOpen
  \bibfield  {author} {\bibinfo {author} {\bibfnamefont {S.}~\bibnamefont
  {Rahav}}\ and\ \bibinfo {author} {\bibfnamefont {C.}~\bibnamefont
  {Jarzynski}},\ }\bibfield  {title} {\bibinfo {title} {Fluctuation relations
  and coarse-graining},\ }\href
  {https://doi.org/10.1088/1742-5468/2007/09/P09012} {\bibfield  {journal}
  {\bibinfo  {journal} {Journal of Statistical Mechanics: Theory and
  Experiment}\ }\textbf {\bibinfo {volume} {2007}},\ \bibinfo {pages} {P09012}
  (\bibinfo {year} {2007})}\BibitemShut {NoStop}%
\bibitem [{\citenamefont {Hartich}\ and\ \citenamefont
  {Godec}(2021{\natexlab{a}})}]{Godec_KineticHysteresis}%
  \BibitemOpen
  \bibfield  {author} {\bibinfo {author} {\bibfnamefont {D.}~\bibnamefont
  {Hartich}}\ and\ \bibinfo {author} {\bibfnamefont {A.~c.~v.}\ \bibnamefont
  {Godec}},\ }\bibfield  {title} {\bibinfo {title} {Emergent memory and kinetic
  hysteresis in strongly driven networks},\ }\href
  {https://doi.org/10.1103/PhysRevX.11.041047} {\bibfield  {journal} {\bibinfo
  {journal} {Phys. Rev. X}\ }\textbf {\bibinfo {volume} {11}},\ \bibinfo
  {pages} {041047} (\bibinfo {year} {2021}{\natexlab{a}})}\BibitemShut
  {NoStop}%
\bibitem [{\citenamefont {Esposito}(2012)}]{Esposito_CG_2012}%
  \BibitemOpen
  \bibfield  {author} {\bibinfo {author} {\bibfnamefont {M.}~\bibnamefont
  {Esposito}},\ }\bibfield  {title} {\bibinfo {title} {Stochastic
  thermodynamics under coarse graining},\ }\href
  {https://doi.org/10.1103/PhysRevE.85.041125} {\bibfield  {journal} {\bibinfo
  {journal} {Phys. Rev. E}\ }\textbf {\bibinfo {volume} {85}},\ \bibinfo
  {pages} {041125} (\bibinfo {year} {2012})}\BibitemShut {NoStop}%
\bibitem [{\citenamefont {Kawai}\ \emph {et~al.}(2007)\citenamefont {Kawai},
  \citenamefont {Parrondo},\ and\ \citenamefont {den
  Broeck}}]{Kawai_DissipationPhaseSpace}%
  \BibitemOpen
  \bibfield  {author} {\bibinfo {author} {\bibfnamefont {R.}~\bibnamefont
  {Kawai}}, \bibinfo {author} {\bibfnamefont {J.~M.~R.}\ \bibnamefont
  {Parrondo}},\ and\ \bibinfo {author} {\bibfnamefont {C.~V.}\ \bibnamefont
  {den Broeck}},\ }\bibfield  {title} {\bibinfo {title} {Dissipation: The
  phase-space perspective},\ }\href
  {https://doi.org/10.1103/PhysRevLett.98.080602} {\bibfield  {journal}
  {\bibinfo  {journal} {Phys. Rev. Lett.}\ }\textbf {\bibinfo {volume} {98}},\
  \bibinfo {pages} {080602} (\bibinfo {year} {2007})}\BibitemShut {NoStop}%
\bibitem [{\citenamefont {Li}\ \emph {et~al.}(2019)\citenamefont {Li},
  \citenamefont {Horowitz}, \citenamefont {Gingrich},\ and\ \citenamefont
  {Fakhri}}]{Fakhri_Jordan_2019}%
  \BibitemOpen
  \bibfield  {author} {\bibinfo {author} {\bibfnamefont {J.}~\bibnamefont
  {Li}}, \bibinfo {author} {\bibfnamefont {J.~M.}\ \bibnamefont {Horowitz}},
  \bibinfo {author} {\bibfnamefont {T.~R.}\ \bibnamefont {Gingrich}},\ and\
  \bibinfo {author} {\bibfnamefont {N.}~\bibnamefont {Fakhri}},\ }\bibfield
  {title} {\bibinfo {title} {Quantifying dissipation using fluctuating
  currents},\ }\href {https://doi.org/10.1038/s41467-019-09631-x} {\bibfield
  {journal} {\bibinfo  {journal} {Nature Communications}\ }\textbf {\bibinfo
  {volume} {10}},\ \bibinfo {pages} {1666} (\bibinfo {year}
  {2019})}\BibitemShut {NoStop}%
\bibitem [{\citenamefont {Bisker}\ \emph {et~al.}(2017)\citenamefont {Bisker},
  \citenamefont {Polettini}, \citenamefont {Gingrich},\ and\ \citenamefont
  {Horowitz}}]{Bisker_2017}%
  \BibitemOpen
  \bibfield  {author} {\bibinfo {author} {\bibfnamefont {G.}~\bibnamefont
  {Bisker}}, \bibinfo {author} {\bibfnamefont {M.}~\bibnamefont {Polettini}},
  \bibinfo {author} {\bibfnamefont {T.~R.}\ \bibnamefont {Gingrich}},\ and\
  \bibinfo {author} {\bibfnamefont {J.~M.}\ \bibnamefont {Horowitz}},\
  }\bibfield  {title} {\bibinfo {title} {Hierarchical bounds on entropy
  production inferred from partial information},\ }\href
  {https://doi.org/10.1088/1742-5468/aa8c0d} {\bibfield  {journal} {\bibinfo
  {journal} {Journal of Statistical Mechanics: Theory and Experiment}\ }\textbf
  {\bibinfo {volume} {2017}},\ \bibinfo {pages} {093210} (\bibinfo {year}
  {2017})}\BibitemShut {NoStop}%
\bibitem [{\citenamefont {Seiferth}\ \emph {et~al.}(2020)\citenamefont
  {Seiferth}, \citenamefont {Sollich},\ and\ \citenamefont
  {Klumpp}}]{CGexamples}%
  \BibitemOpen
  \bibfield  {author} {\bibinfo {author} {\bibfnamefont {D.}~\bibnamefont
  {Seiferth}}, \bibinfo {author} {\bibfnamefont {P.}~\bibnamefont {Sollich}},\
  and\ \bibinfo {author} {\bibfnamefont {S.}~\bibnamefont {Klumpp}},\
  }\bibfield  {title} {\bibinfo {title} {Coarse graining of biochemical systems
  described by discrete stochastic dynamics},\ }\href
  {https://doi.org/10.1103/PhysRevE.102.062149} {\bibfield  {journal} {\bibinfo
   {journal} {Phys. Rev. E}\ }\textbf {\bibinfo {volume} {102}},\ \bibinfo
  {pages} {062149} (\bibinfo {year} {2020})}\BibitemShut {NoStop}%
\bibitem [{\citenamefont {Bilotto}\ \emph {et~al.}(2021)\citenamefont
  {Bilotto}, \citenamefont {Caprini},\ and\ \citenamefont
  {Vulpiani}}]{CGexamples2}%
  \BibitemOpen
  \bibfield  {author} {\bibinfo {author} {\bibfnamefont {P.}~\bibnamefont
  {Bilotto}}, \bibinfo {author} {\bibfnamefont {L.}~\bibnamefont {Caprini}},\
  and\ \bibinfo {author} {\bibfnamefont {A.}~\bibnamefont {Vulpiani}},\
  }\bibfield  {title} {\bibinfo {title} {Excess and loss of entropy production
  for different levels of coarse graining},\ }\href
  {https://doi.org/10.1103/PhysRevE.104.024140} {\bibfield  {journal} {\bibinfo
   {journal} {Phys. Rev. E}\ }\textbf {\bibinfo {volume} {104}},\ \bibinfo
  {pages} {024140} (\bibinfo {year} {2021})}\BibitemShut {NoStop}%
\bibitem [{\citenamefont {Puglisi}\ \emph {et~al.}(2010)\citenamefont
  {Puglisi}, \citenamefont {Pigolotti}, \citenamefont {Rondoni},\ and\
  \citenamefont {Vulpiani}}]{EPR_CG}%
  \BibitemOpen
  \bibfield  {author} {\bibinfo {author} {\bibfnamefont {A.}~\bibnamefont
  {Puglisi}}, \bibinfo {author} {\bibfnamefont {S.}~\bibnamefont {Pigolotti}},
  \bibinfo {author} {\bibfnamefont {L.}~\bibnamefont {Rondoni}},\ and\ \bibinfo
  {author} {\bibfnamefont {A.}~\bibnamefont {Vulpiani}},\ }\bibfield  {title}
  {\bibinfo {title} {Entropy production and coarse graining in markov
  processes},\ }\href {https://doi.org/10.1088/1742-5468/2010/05/p05015}
  {\bibfield  {journal} {\bibinfo  {journal} {Journal of Statistical Mechanics:
  Theory and Experiment}\ }\textbf {\bibinfo {volume} {2010}},\ \bibinfo
  {pages} {P05015} (\bibinfo {year} {2010})}\BibitemShut {NoStop}%
\bibitem [{\citenamefont {Teza}\ and\ \citenamefont
  {Stella}(2020)}]{Teza_ExactCG_PhysRevLett.125.110601}%
  \BibitemOpen
  \bibfield  {author} {\bibinfo {author} {\bibfnamefont {G.}~\bibnamefont
  {Teza}}\ and\ \bibinfo {author} {\bibfnamefont {A.~L.}\ \bibnamefont
  {Stella}},\ }\bibfield  {title} {\bibinfo {title} {Exact coarse graining
  preserves entropy production out of equilibrium},\ }\href
  {https://doi.org/10.1103/PhysRevLett.125.110601} {\bibfield  {journal}
  {\bibinfo  {journal} {Phys. Rev. Lett.}\ }\textbf {\bibinfo {volume} {125}},\
  \bibinfo {pages} {110601} (\bibinfo {year} {2020})}\BibitemShut {NoStop}%
\bibitem [{\citenamefont {Harunari}\ \emph {et~al.}(2022)\citenamefont
  {Harunari}, \citenamefont {Dutta}, \citenamefont {Polettini},\ and\
  \citenamefont {Rold{\'a}n}}]{TransitionBased_Roldan}%
  \BibitemOpen
  \bibfield  {author} {\bibinfo {author} {\bibfnamefont {P.~E.}\ \bibnamefont
  {Harunari}}, \bibinfo {author} {\bibfnamefont {A.}~\bibnamefont {Dutta}},
  \bibinfo {author} {\bibfnamefont {M.}~\bibnamefont {Polettini}},\ and\
  \bibinfo {author} {\bibfnamefont {{\'E}.}~\bibnamefont {Rold{\'a}n}},\
  }\bibfield  {title} {\bibinfo {title} {What to learn from few visible
  transitions' statistics?},\ }\Eprint {https://arxiv.org/abs/2203.07427}
  {arXiv:2203.07427}  (\bibinfo {year} {2022})\BibitemShut {NoStop}%
\bibitem [{\citenamefont {van~der Meer}\ \emph
  {et~al.}(2022{\natexlab{a}})\citenamefont {van~der Meer}, \citenamefont
  {Ertel},\ and\ \citenamefont {Seifert}}]{TransitionBased_Seifrit}%
  \BibitemOpen
  \bibfield  {author} {\bibinfo {author} {\bibfnamefont {J.}~\bibnamefont
  {van~der Meer}}, \bibinfo {author} {\bibfnamefont {B.}~\bibnamefont
  {Ertel}},\ and\ \bibinfo {author} {\bibfnamefont {U.}~\bibnamefont
  {Seifert}},\ }\bibfield  {title} {\bibinfo {title} {Thermodynamic inference
  in partially accessible markov networks: A unifying perspective from
  transition-based waiting time distributions},\ }\href
  {https://doi.org/10.1103/PhysRevX.12.031025} {\bibfield  {journal} {\bibinfo
  {journal} {Phys. Rev. X}\ }\textbf {\bibinfo {volume} {12}},\ \bibinfo
  {pages} {031025} (\bibinfo {year} {2022}{\natexlab{a}})}\BibitemShut
  {NoStop}%
\bibitem [{\citenamefont {van~der Meer}\ \emph
  {et~al.}(2022{\natexlab{b}})\citenamefont {van~der Meer}, \citenamefont
  {Degünther},\ and\ \citenamefont {Seifert}}]{Seifrit2022}%
  \BibitemOpen
  \bibfield  {author} {\bibinfo {author} {\bibfnamefont {J.}~\bibnamefont
  {van~der Meer}}, \bibinfo {author} {\bibfnamefont {J.}~\bibnamefont
  {Degünther}},\ and\ \bibinfo {author} {\bibfnamefont {U.}~\bibnamefont
  {Seifert}},\ }\href@noop {} {\bibinfo {title} {Time-resolved statistics of
  snippets as general framework for model-free entropy estimators}} (\bibinfo
  {year} {2022}{\natexlab{b}}),\ \Eprint {https://arxiv.org/abs/2211.17032}
  {arXiv:2211.17032} \BibitemShut {NoStop}%
\bibitem [{\citenamefont
  {Shiraishi}(2021{\natexlab{a}})}]{shiraishi2021optimal}%
  \BibitemOpen
  \bibfield  {author} {\bibinfo {author} {\bibfnamefont {N.}~\bibnamefont
  {Shiraishi}},\ }\bibfield  {title} {\bibinfo {title} {Optimal thermodynamic
  uncertainty relation in markov jump processes},\ }\href
  {https://doi.org/https://doi.org/10.1007/s10955-021-02829-8} {\bibfield
  {journal} {\bibinfo  {journal} {Journal of Statistical Physics}\ }\textbf
  {\bibinfo {volume} {185}},\ \bibinfo {pages} {1} (\bibinfo {year}
  {2021}{\natexlab{a}})}\BibitemShut {NoStop}%
\bibitem [{\citenamefont {Horowitz}\ and\ \citenamefont
  {Gingrich}(2020)}]{horowitz2020thermodynamic}%
  \BibitemOpen
  \bibfield  {author} {\bibinfo {author} {\bibfnamefont {J.~M.}\ \bibnamefont
  {Horowitz}}\ and\ \bibinfo {author} {\bibfnamefont {T.~R.}\ \bibnamefont
  {Gingrich}},\ }\bibfield  {title} {\bibinfo {title} {Thermodynamic
  uncertainty relations constrain non-equilibrium fluctuations},\ }\href
  {https://doi.org/https://doi.org/10.1038/s41567-019-0702-6} {\bibfield
  {journal} {\bibinfo  {journal} {Nature Physics}\ }\textbf {\bibinfo {volume}
  {16}},\ \bibinfo {pages} {15} (\bibinfo {year} {2020})}\BibitemShut {NoStop}%
\bibitem [{\citenamefont {Falasco}\ \emph {et~al.}(2020)\citenamefont
  {Falasco}, \citenamefont {Esposito},\ and\ \citenamefont
  {Delvenne}}]{falasco2020unifying}%
  \BibitemOpen
  \bibfield  {author} {\bibinfo {author} {\bibfnamefont {G.}~\bibnamefont
  {Falasco}}, \bibinfo {author} {\bibfnamefont {M.}~\bibnamefont {Esposito}},\
  and\ \bibinfo {author} {\bibfnamefont {J.-C.}\ \bibnamefont {Delvenne}},\
  }\bibfield  {title} {\bibinfo {title} {Unifying thermodynamic uncertainty
  relations},\ }\href {https://doi.org/10.1088/1367-2630/ab8679} {\bibfield
  {journal} {\bibinfo  {journal} {New Journal of Physics}\ }\textbf {\bibinfo
  {volume} {22}},\ \bibinfo {pages} {053046} (\bibinfo {year}
  {2020})}\BibitemShut {NoStop}%
\bibitem [{\citenamefont {Barato}\ and\ \citenamefont
  {Seifert}(2015)}]{barato2015thermodynamic}%
  \BibitemOpen
  \bibfield  {author} {\bibinfo {author} {\bibfnamefont {A.~C.}\ \bibnamefont
  {Barato}}\ and\ \bibinfo {author} {\bibfnamefont {U.}~\bibnamefont
  {Seifert}},\ }\bibfield  {title} {\bibinfo {title} {Thermodynamic uncertainty
  relation for biomolecular processes},\ }\href
  {https://doi.org/10.1103/PhysRevLett.114.158101} {\bibfield  {journal}
  {\bibinfo  {journal} {Physical review letters}\ }\textbf {\bibinfo {volume}
  {114}},\ \bibinfo {pages} {158101} (\bibinfo {year} {2015})}\BibitemShut
  {NoStop}%
\bibitem [{\citenamefont {Otsubo}\ \emph {et~al.}(2020)\citenamefont {Otsubo},
  \citenamefont {Ito}, \citenamefont {Dechant},\ and\ \citenamefont
  {Sagawa}}]{TUR_ML_Sagawa}%
  \BibitemOpen
  \bibfield  {author} {\bibinfo {author} {\bibfnamefont {S.}~\bibnamefont
  {Otsubo}}, \bibinfo {author} {\bibfnamefont {S.}~\bibnamefont {Ito}},
  \bibinfo {author} {\bibfnamefont {A.}~\bibnamefont {Dechant}},\ and\ \bibinfo
  {author} {\bibfnamefont {T.}~\bibnamefont {Sagawa}},\ }\bibfield  {title}
  {\bibinfo {title} {Estimating entropy production by machine learning of
  short-time fluctuating currents},\ }\href
  {https://doi.org/10.1103/PhysRevE.101.062106} {\bibfield  {journal} {\bibinfo
   {journal} {Phys. Rev. E}\ }\textbf {\bibinfo {volume} {101}},\ \bibinfo
  {pages} {062106} (\bibinfo {year} {2020})}\BibitemShut {NoStop}%
\bibitem [{\citenamefont {Seifert}(2019)}]{TUR_seifrit2}%
  \BibitemOpen
  \bibfield  {author} {\bibinfo {author} {\bibfnamefont {U.}~\bibnamefont
  {Seifert}},\ }\bibfield  {title} {\bibinfo {title} {From stochastic
  thermodynamics to thermodynamic inference},\ }\href
  {https://doi.org/10.1146/annurev-conmatphys-031218-013554} {\bibfield
  {journal} {\bibinfo  {journal} {Annual Review of Condensed Matter Physics}\
  }\textbf {\bibinfo {volume} {10}},\ \bibinfo {pages} {171} (\bibinfo {year}
  {2019})}\BibitemShut {NoStop}%
\bibitem [{\citenamefont {Vu}\ and\ \citenamefont
  {Hasegawa}(2020{\natexlab{a}})}]{TURforSemiMarkov_Hasegawa}%
  \BibitemOpen
  \bibfield  {author} {\bibinfo {author} {\bibfnamefont {T.~V.}\ \bibnamefont
  {Vu}}\ and\ \bibinfo {author} {\bibfnamefont {Y.}~\bibnamefont {Hasegawa}},\
  }\bibfield  {title} {\bibinfo {title} {Generalized uncertainty relations for
  semi-markov processes},\ }\href
  {https://doi.org/10.1088/1742-6596/1593/1/012006} {\bibfield  {journal}
  {\bibinfo  {journal} {Journal of Physics: Conference Series}\ }\textbf
  {\bibinfo {volume} {1593}},\ \bibinfo {pages} {012006} (\bibinfo {year}
  {2020}{\natexlab{a}})}\BibitemShut {NoStop}%
\bibitem [{\citenamefont {Ertel}\ \emph {et~al.}(2022)\citenamefont {Ertel},
  \citenamefont {van~der Meer},\ and\ \citenamefont
  {Seifert}}]{TURforSemiMarkov_Seifrit}%
  \BibitemOpen
  \bibfield  {author} {\bibinfo {author} {\bibfnamefont {B.}~\bibnamefont
  {Ertel}}, \bibinfo {author} {\bibfnamefont {J.}~\bibnamefont {van~der
  Meer}},\ and\ \bibinfo {author} {\bibfnamefont {U.}~\bibnamefont {Seifert}},\
  }\bibfield  {title} {\bibinfo {title} {Operationally accessible uncertainty
  relations for thermodynamically consistent semi-markov processes},\ }\href
  {https://doi.org/10.1103/PhysRevE.105.044113} {\bibfield  {journal} {\bibinfo
   {journal} {Phys. Rev. E}\ }\textbf {\bibinfo {volume} {105}},\ \bibinfo
  {pages} {044113} (\bibinfo {year} {2022})}\BibitemShut {NoStop}%
\bibitem [{\citenamefont {Kamijima}\ \emph {et~al.}(2021)\citenamefont
  {Kamijima}, \citenamefont {Otsubo}, \citenamefont {Ashida},\ and\
  \citenamefont {Sagawa}}]{TURderivation_Sagawa}%
  \BibitemOpen
  \bibfield  {author} {\bibinfo {author} {\bibfnamefont {T.}~\bibnamefont
  {Kamijima}}, \bibinfo {author} {\bibfnamefont {S.}~\bibnamefont {Otsubo}},
  \bibinfo {author} {\bibfnamefont {Y.}~\bibnamefont {Ashida}},\ and\ \bibinfo
  {author} {\bibfnamefont {T.}~\bibnamefont {Sagawa}},\ }\bibfield  {title}
  {\bibinfo {title} {Higher-order efficiency bound and its application to
  nonlinear nanothermoelectrics},\ }\href
  {https://doi.org/10.1103/PhysRevE.104.044115} {\bibfield  {journal} {\bibinfo
   {journal} {Phys. Rev. E}\ }\textbf {\bibinfo {volume} {104}},\ \bibinfo
  {pages} {044115} (\bibinfo {year} {2021})}\BibitemShut {NoStop}%
\bibitem [{\citenamefont
  {Shiraishi}(2021{\natexlab{b}})}]{TURderivation_Shiraishi}%
  \BibitemOpen
  \bibfield  {author} {\bibinfo {author} {\bibfnamefont {N.}~\bibnamefont
  {Shiraishi}},\ }\bibfield  {title} {\bibinfo {title} {Inferring broken
  detailed balance in the absence of observable currents.},\ }\href
  {https://doi.org/10.1007/s10955-021-02829-8} {\bibfield  {journal} {\bibinfo
  {journal} {J. Stat. Phys.}\ }\textbf {\bibinfo {volume} {185}},\ \bibinfo
  {pages} {5} (\bibinfo {year} {2021}{\natexlab{b}})}\BibitemShut {NoStop}%
\bibitem [{\citenamefont {K~Manikandan}\ \emph {et~al.}(2021)\citenamefont
  {K~Manikandan}, \citenamefont {Ghosh}, \citenamefont {Kundu}, \citenamefont
  {Das}, \citenamefont {Agrawal}, \citenamefont {Mitra}, \citenamefont
  {Banerjee},\ and\ \citenamefont {Krishnamurthy}}]{TURnature}%
  \BibitemOpen
  \bibfield  {author} {\bibinfo {author} {\bibfnamefont {S.}~\bibnamefont
  {K~Manikandan}}, \bibinfo {author} {\bibfnamefont {S.}~\bibnamefont {Ghosh}},
  \bibinfo {author} {\bibfnamefont {A.}~\bibnamefont {Kundu}}, \bibinfo
  {author} {\bibfnamefont {B.}~\bibnamefont {Das}}, \bibinfo {author}
  {\bibfnamefont {V.}~\bibnamefont {Agrawal}}, \bibinfo {author} {\bibfnamefont
  {D.}~\bibnamefont {Mitra}}, \bibinfo {author} {\bibfnamefont
  {A.}~\bibnamefont {Banerjee}},\ and\ \bibinfo {author} {\bibfnamefont
  {S.}~\bibnamefont {Krishnamurthy}},\ }\bibfield  {title} {\bibinfo {title}
  {Quantitative analysis of non-equilibrium systems from short-time
  experimental data},\ }\href {https://doi.org/10.1038/s42005-021-00766-2}
  {\bibfield  {journal} {\bibinfo  {journal} {Communications Physics}\ }\textbf
  {\bibinfo {volume} {4}} (\bibinfo {year} {2021})}\BibitemShut {NoStop}%
\bibitem [{\citenamefont {Vu}\ and\ \citenamefont
  {Hasegawa}(2020{\natexlab{b}})}]{GeneralizedTUR}%
  \BibitemOpen
  \bibfield  {author} {\bibinfo {author} {\bibfnamefont {T.~V.}\ \bibnamefont
  {Vu}}\ and\ \bibinfo {author} {\bibfnamefont {Y.}~\bibnamefont {Hasegawa}},\
  }\bibfield  {title} {\bibinfo {title} {Generalized uncertainty relations for
  semi-markov processes},\ }\href
  {https://doi.org/10.1088/1742-6596/1593/1/012006} {\bibfield  {journal}
  {\bibinfo  {journal} {Journal of Physics: Conference Series}\ }\textbf
  {\bibinfo {volume} {1593}},\ \bibinfo {pages} {012006} (\bibinfo {year}
  {2020}{\natexlab{b}})}\BibitemShut {NoStop}%
\bibitem [{\citenamefont {Skinner}\ and\ \citenamefont
  {Dunkel}(2021{\natexlab{a}})}]{Dunkel_OPT}%
  \BibitemOpen
  \bibfield  {author} {\bibinfo {author} {\bibfnamefont {D.~J.}\ \bibnamefont
  {Skinner}}\ and\ \bibinfo {author} {\bibfnamefont {J.}~\bibnamefont
  {Dunkel}},\ }\bibfield  {title} {\bibinfo {title} {Improved bounds on entropy
  production in living systems},\ }\href
  {https://doi.org/10.1073/pnas.2024300118} {\bibfield  {journal} {\bibinfo
  {journal} {Proceedings of the National Academy of Sciences}\ }\textbf
  {\bibinfo {volume} {118}},\ \bibinfo {pages} {e2024300118} (\bibinfo {year}
  {2021}{\natexlab{a}})}\BibitemShut {NoStop}%
\bibitem [{\citenamefont {Skinner}\ and\ \citenamefont
  {Dunkel}(2021{\natexlab{b}})}]{Dunkel_WTD}%
  \BibitemOpen
  \bibfield  {author} {\bibinfo {author} {\bibfnamefont {D.~J.}\ \bibnamefont
  {Skinner}}\ and\ \bibinfo {author} {\bibfnamefont {J.}~\bibnamefont
  {Dunkel}},\ }\bibfield  {title} {\bibinfo {title} {Estimating entropy
  production from waiting time distributions},\ }\href
  {https://doi.org/10.1103/PhysRevLett.127.198101} {\bibfield  {journal}
  {\bibinfo  {journal} {Phys. Rev. Lett.}\ }\textbf {\bibinfo {volume} {127}},\
  \bibinfo {pages} {198101} (\bibinfo {year} {2021}{\natexlab{b}})}\BibitemShut
  {NoStop}%
\bibitem [{\citenamefont {Nitzan}\ \emph {et~al.}(2022)\citenamefont {Nitzan},
  \citenamefont {Ghosal},\ and\ \citenamefont {Bisker}}]{Eden2022}%
  \BibitemOpen
  \bibfield  {author} {\bibinfo {author} {\bibfnamefont {E.}~\bibnamefont
  {Nitzan}}, \bibinfo {author} {\bibfnamefont {A.}~\bibnamefont {Ghosal}},\
  and\ \bibinfo {author} {\bibfnamefont {G.}~\bibnamefont {Bisker}},\
  }\href@noop {} {\bibinfo {title} {Universal bounds on entropy production
  inferred from observed statistics}} (\bibinfo {year} {2022}),\ \Eprint
  {https://arxiv.org/abs/2212.01783} {arXiv:2212.01783} \BibitemShut {NoStop}%
\bibitem [{\citenamefont {Ehrich}(2021)}]{AprioriEstimator}%
  \BibitemOpen
  \bibfield  {author} {\bibinfo {author} {\bibfnamefont {J.}~\bibnamefont
  {Ehrich}},\ }\bibfield  {title} {\bibinfo {title} {Tightest bound on hidden
  entropy production from partially observed dynamics},\ }\href
  {https://doi.org/10.1088/1742-5468/ac150e} {\bibfield  {journal} {\bibinfo
  {journal} {Journal of Statistical Mechanics: Theory and Experiment}\ }\textbf
  {\bibinfo {volume} {2021}},\ \bibinfo {pages} {083214} (\bibinfo {year}
  {2021})}\BibitemShut {NoStop}%
\bibitem [{\citenamefont {Parrondo}\ \emph {et~al.}(2009)\citenamefont
  {Parrondo}, \citenamefont {Broeck},\ and\ \citenamefont
  {Kawai}}]{parrondo2009arrow}%
  \BibitemOpen
  \bibfield  {author} {\bibinfo {author} {\bibfnamefont {J.~M.}\ \bibnamefont
  {Parrondo}}, \bibinfo {author} {\bibfnamefont {C.}~\bibnamefont {Broeck}},\
  and\ \bibinfo {author} {\bibfnamefont {R.}~\bibnamefont {Kawai}},\ }\bibfield
   {title} {\bibinfo {title} {Entropy production and the arrow of time},\
  }\href {https://doi.org/10.1088/1367-2630/11/7/073008} {\bibfield  {journal}
  {\bibinfo  {journal} {New Journal of Physics}\ }\textbf {\bibinfo {volume}
  {11}} (\bibinfo {year} {2009})}\BibitemShut {NoStop}%
\bibitem [{\citenamefont {Mart´ınez}\ \emph {et~al.}(2019)\citenamefont
  {Mart´ınez}, \citenamefont {Bisker}, \citenamefont {Horowitz},\ and\
  \citenamefont {Parrondo}}]{GiliNat19}%
  \BibitemOpen
  \bibfield  {author} {\bibinfo {author} {\bibfnamefont {I.~A.}\ \bibnamefont
  {Mart´ınez}}, \bibinfo {author} {\bibfnamefont {G.}~\bibnamefont {Bisker}},
  \bibinfo {author} {\bibfnamefont {J.~M.}\ \bibnamefont {Horowitz}},\ and\
  \bibinfo {author} {\bibfnamefont {J.~M.~R.}\ \bibnamefont {Parrondo}},\
  }\bibfield  {title} {\bibinfo {title} {Inferring broken detailed balance in
  the absence of observable currents.},\ }\href
  {https://doi.org/10.1038/s41467-019-11051-w} {\bibfield  {journal} {\bibinfo
  {journal} {Nat. Commun.}\ }\textbf {\bibinfo {volume} {10}},\ \bibinfo
  {pages} {3542} (\bibinfo {year} {2019})}\BibitemShut {NoStop}%
\bibitem [{\citenamefont {Rached}\ \emph {et~al.}(2004)\citenamefont {Rached},
  \citenamefont {Alajaji},\ and\ \citenamefont {Campbell}}]{KldCalcInMarkov}%
  \BibitemOpen
  \bibfield  {author} {\bibinfo {author} {\bibfnamefont {Z.}~\bibnamefont
  {Rached}}, \bibinfo {author} {\bibfnamefont {F.}~\bibnamefont {Alajaji}},\
  and\ \bibinfo {author} {\bibfnamefont {L.}~\bibnamefont {Campbell}},\
  }\bibfield  {title} {\bibinfo {title} {The kullback-leibler divergence rate
  between markov sources},\ }\href {https://doi.org/10.1109/TIT.2004.826687}
  {\bibfield  {journal} {\bibinfo  {journal} {IEEE Transactions on Information
  Theory}\ }\textbf {\bibinfo {volume} {50}},\ \bibinfo {pages} {917} (\bibinfo
  {year} {2004})}\BibitemShut {NoStop}%
\bibitem [{\citenamefont {Rold\'an}\ and\ \citenamefont
  {Parrondo}(2012)}]{FR_semiCG_semiAnalytical}%
  \BibitemOpen
  \bibfield  {author} {\bibinfo {author} {\bibfnamefont {E.}~\bibnamefont
  {Rold\'an}}\ and\ \bibinfo {author} {\bibfnamefont {J.~M.~R.}\ \bibnamefont
  {Parrondo}},\ }\bibfield  {title} {\bibinfo {title} {Entropy production and
  kullback-leibler divergence between stationary trajectories of discrete
  systems},\ }\href {https://doi.org/10.1103/PhysRevE.85.031129} {\bibfield
  {journal} {\bibinfo  {journal} {Phys. Rev. E}\ }\textbf {\bibinfo {volume}
  {85}},\ \bibinfo {pages} {031129} (\bibinfo {year} {2012})}\BibitemShut
  {NoStop}%
\bibitem [{\citenamefont {Rold\'an}\ and\ \citenamefont
  {Parrondo}(2010)}]{roldanFR}%
  \BibitemOpen
  \bibfield  {author} {\bibinfo {author} {\bibfnamefont {E.}~\bibnamefont
  {Rold\'an}}\ and\ \bibinfo {author} {\bibfnamefont {J.~M.~R.}\ \bibnamefont
  {Parrondo}},\ }\bibfield  {title} {\bibinfo {title} {Estimating dissipation
  from single stationary trajectories},\ }\href
  {https://doi.org/10.1103/PhysRevLett.105.150607} {\bibfield  {journal}
  {\bibinfo  {journal} {Phys. Rev. Lett.}\ }\textbf {\bibinfo {volume} {105}},\
  \bibinfo {pages} {150607} (\bibinfo {year} {2010})}\BibitemShut {NoStop}%
\bibitem [{\citenamefont {Ghosal}\ and\ \citenamefont
  {Bisker}(2022)}]{AishaniEPR}%
  \BibitemOpen
  \bibfield  {author} {\bibinfo {author} {\bibfnamefont {A.}~\bibnamefont
  {Ghosal}}\ and\ \bibinfo {author} {\bibfnamefont {G.}~\bibnamefont
  {Bisker}},\ }\bibfield  {title} {\bibinfo {title} {Inferring entropy
  production rate from partially observed langevin dynamics under
  coarse-graining},\ }\href {https://doi.org/10.1039/D2CP03064K} {\bibfield
  {journal} {\bibinfo  {journal} {Physical Chemistry Chemical Physics}\ }
  (\bibinfo {year} {2022})}\BibitemShut {NoStop}%
\bibitem [{\citenamefont {Ro}\ \emph {et~al.}(2022)\citenamefont {Ro},
  \citenamefont {Guo}, \citenamefont {Shih}, \citenamefont {Phan},
  \citenamefont {Austin}, \citenamefont {Levine}, \citenamefont {Chaikin},\
  and\ \citenamefont {Martiniani}}]{Ro_PhysRevLett.129.220601_Active_Matter}%
  \BibitemOpen
  \bibfield  {author} {\bibinfo {author} {\bibfnamefont {S.}~\bibnamefont
  {Ro}}, \bibinfo {author} {\bibfnamefont {B.}~\bibnamefont {Guo}}, \bibinfo
  {author} {\bibfnamefont {A.}~\bibnamefont {Shih}}, \bibinfo {author}
  {\bibfnamefont {T.~V.}\ \bibnamefont {Phan}}, \bibinfo {author}
  {\bibfnamefont {R.~H.}\ \bibnamefont {Austin}}, \bibinfo {author}
  {\bibfnamefont {D.}~\bibnamefont {Levine}}, \bibinfo {author} {\bibfnamefont
  {P.~M.}\ \bibnamefont {Chaikin}},\ and\ \bibinfo {author} {\bibfnamefont
  {S.}~\bibnamefont {Martiniani}},\ }\bibfield  {title} {\bibinfo {title}
  {Model-free measurement of local entropy production and extractable work in
  active matter},\ }\href {https://doi.org/10.1103/PhysRevLett.129.220601}
  {\bibfield  {journal} {\bibinfo  {journal} {Phys. Rev. Lett.}\ }\textbf
  {\bibinfo {volume} {129}},\ \bibinfo {pages} {220601} (\bibinfo {year}
  {2022})}\BibitemShut {NoStop}%
\bibitem [{\citenamefont {Wang}\ \emph {et~al.}(2005)\citenamefont {Wang},
  \citenamefont {Kulkarni},\ and\ \citenamefont
  {Verd{\'u}}}]{Wang2005DivergenceEO}%
  \BibitemOpen
  \bibfield  {author} {\bibinfo {author} {\bibfnamefont {Q.}~\bibnamefont
  {Wang}}, \bibinfo {author} {\bibfnamefont {S.~R.}\ \bibnamefont {Kulkarni}},\
  and\ \bibinfo {author} {\bibfnamefont {S.}~\bibnamefont {Verd{\'u}}},\
  }\bibfield  {title} {\bibinfo {title} {Divergence estimation of continuous
  distributions based on data-dependent partitions},\ }\href
  {https://doi.org/10.1109/TIT.2005.853314} {\bibfield  {journal} {\bibinfo
  {journal} {IEEE Transactions on Information Theory}\ }\textbf {\bibinfo
  {volume} {51}},\ \bibinfo {pages} {3064} (\bibinfo {year}
  {2005})}\BibitemShut {NoStop}%
\bibitem [{\citenamefont {Bisker}\ \emph {et~al.}(2022)\citenamefont {Bisker},
  \citenamefont {Martinez}, \citenamefont {Horowitz},\ and\ \citenamefont
  {Parrondo}}]{GiliCommentonGodec}%
  \BibitemOpen
  \bibfield  {author} {\bibinfo {author} {\bibfnamefont {G.}~\bibnamefont
  {Bisker}}, \bibinfo {author} {\bibfnamefont {I.~A.}\ \bibnamefont
  {Martinez}}, \bibinfo {author} {\bibfnamefont {J.~M.}\ \bibnamefont
  {Horowitz}},\ and\ \bibinfo {author} {\bibfnamefont {J.~M.}\ \bibnamefont
  {Parrondo}},\ }\href@noop {} {\bibinfo {title} {Comment on "inferring broken
  detailed balance in the absence of observable currents"}} (\bibinfo {year}
  {2022}),\ \Eprint {https://arxiv.org/abs/2202.02064} {arXiv:2202.02064}
  \BibitemShut {NoStop}%
\bibitem [{\citenamefont {Hartich}\ and\ \citenamefont
  {Godec}(2021{\natexlab{b}})}]{GodecCommentonGili}%
  \BibitemOpen
  \bibfield  {author} {\bibinfo {author} {\bibfnamefont {D.}~\bibnamefont
  {Hartich}}\ and\ \bibinfo {author} {\bibfnamefont {A.}~\bibnamefont
  {Godec}},\ }\href@noop {} {\bibinfo {title} {Comment on "inferring broken
  detailed balance in the absence of observable currents"}} (\bibinfo {year}
  {2021}{\natexlab{b}}),\ \Eprint {https://arxiv.org/abs/2112.08978}
  {arXiv:2112.08978} \BibitemShut {NoStop}%
\bibitem [{\citenamefont {Kim}\ \emph {et~al.}(2020)\citenamefont {Kim},
  \citenamefont {Bae}, \citenamefont {Lee},\ and\ \citenamefont
  {Jeong}}]{NEEP}%
  \BibitemOpen
  \bibfield  {author} {\bibinfo {author} {\bibfnamefont {D.-K.}\ \bibnamefont
  {Kim}}, \bibinfo {author} {\bibfnamefont {Y.}~\bibnamefont {Bae}}, \bibinfo
  {author} {\bibfnamefont {S.}~\bibnamefont {Lee}},\ and\ \bibinfo {author}
  {\bibfnamefont {H.}~\bibnamefont {Jeong}},\ }\bibfield  {title} {\bibinfo
  {title} {Learning entropy production via neural networks},\ }\href
  {https://doi.org/10.1103/PhysRevLett.125.140604} {\bibfield  {journal}
  {\bibinfo  {journal} {Phys. Rev. Lett.}\ }\textbf {\bibinfo {volume} {125}},\
  \bibinfo {pages} {140604} (\bibinfo {year} {2020})}\BibitemShut {NoStop}%
\bibitem [{\citenamefont {Bae}\ \emph {et~al.}(2022)\citenamefont {Bae},
  \citenamefont {Kim},\ and\ \citenamefont {Jeong}}]{NEEP2}%
  \BibitemOpen
  \bibfield  {author} {\bibinfo {author} {\bibfnamefont {Y.}~\bibnamefont
  {Bae}}, \bibinfo {author} {\bibfnamefont {D.-K.}\ \bibnamefont {Kim}},\ and\
  \bibinfo {author} {\bibfnamefont {H.}~\bibnamefont {Jeong}},\ }\bibfield
  {title} {\bibinfo {title} {Inferring dissipation maps from videos using
  convolutional neural networks},\ }\href
  {https://doi.org/10.1103/PhysRevResearch.4.033094} {\bibfield  {journal}
  {\bibinfo  {journal} {Phys. Rev. Research}\ }\textbf {\bibinfo {volume}
  {4}},\ \bibinfo {pages} {033094} (\bibinfo {year} {2022})}\BibitemShut
  {NoStop}%
\bibitem [{\citenamefont {Otsubo}\ \emph
  {et~al.}(2022{\natexlab{a}})\citenamefont {Otsubo}, \citenamefont
  {Manikandan},\ and\ \citenamefont {Sagawa}}]{NEEP3_Sagawa}%
  \BibitemOpen
  \bibfield  {author} {\bibinfo {author} {\bibfnamefont {S.}~\bibnamefont
  {Otsubo}}, \bibinfo {author} {\bibfnamefont {S.}~\bibnamefont {Manikandan}},\
  and\ \bibinfo {author} {\bibfnamefont {T.}~\bibnamefont {Sagawa}},\
  }\bibfield  {title} {\bibinfo {title} {Estimate non-equilibrium
  trajectories.},\ }\href {https://doi.org/10.1038/s42005-021-00787-x}
  {\bibfield  {journal} {\bibinfo  {journal} {Commun. Phys.}\ }\textbf
  {\bibinfo {volume} {5}},\ \bibinfo {pages} {11} (\bibinfo {year}
  {2022}{\natexlab{a}})}\BibitemShut {NoStop}%
\bibitem [{\citenamefont {Otsubo}\ \emph
  {et~al.}(2022{\natexlab{b}})\citenamefont {Otsubo}, \citenamefont
  {K~Manikandan}, \citenamefont {Sagawa},\ and\ \citenamefont
  {Krishnamurthy}}]{EPR_traj}%
  \BibitemOpen
  \bibfield  {author} {\bibinfo {author} {\bibfnamefont {S.}~\bibnamefont
  {Otsubo}}, \bibinfo {author} {\bibfnamefont {S.}~\bibnamefont
  {K~Manikandan}}, \bibinfo {author} {\bibfnamefont {T.}~\bibnamefont
  {Sagawa}},\ and\ \bibinfo {author} {\bibfnamefont {S.}~\bibnamefont
  {Krishnamurthy}},\ }\bibfield  {title} {\bibinfo {title} {Estimating
  time-dependent entropy production from non-equilibrium trajectories},\ }\href
  {https://doi.org/10.1038/s42005-021-00787-x} {\bibfield  {journal} {\bibinfo
  {journal} {Communications Physics}\ }\textbf {\bibinfo {volume} {5}}
  (\bibinfo {year} {2022}{\natexlab{b}})}\BibitemShut {NoStop}%
\bibitem [{\citenamefont {Schürmann}\ and\ \citenamefont
  {Grassberger}(1996)}]{pluginAnszats}%
  \BibitemOpen
  \bibfield  {author} {\bibinfo {author} {\bibfnamefont {T.}~\bibnamefont
  {Schürmann}}\ and\ \bibinfo {author} {\bibfnamefont {P.}~\bibnamefont
  {Grassberger}},\ }\bibfield  {title} {\bibinfo {title} {Entropy estimation of
  symbol sequences},\ }\href {https://doi.org/10.1063/1.166191} {\bibfield
  {journal} {\bibinfo  {journal} {Chaos: An Interdisciplinary Journal of
  Nonlinear Science}\ }\textbf {\bibinfo {volume} {6}},\ \bibinfo {pages} {414}
  (\bibinfo {year} {1996})}\BibitemShut {NoStop}%
\bibitem [{SI()}]{SI}%
  \BibitemOpen
  \href@noop {} {\bibinfo {title} {See supplemental material at
  \url{https://www.overleaf.com/project/624a88871bc0a5357ca1be9e} for (1)
  waiting-time distributions estimation; (2) plug-in estimator implementation
  details; and (3) rneep convergence in 4-state system.}}\BibitemShut {Stop}%
\bibitem [{\citenamefont {Gillespie}(1977)}]{Gillespie}%
  \BibitemOpen
  \bibfield  {author} {\bibinfo {author} {\bibfnamefont {D.~T.}\ \bibnamefont
  {Gillespie}},\ }\bibfield  {title} {\bibinfo {title} {Exact stochastic
  simulation of coupled chemical reactions},\ }\href
  {https://doi.org/10.1021/j100540a008} {\bibfield  {journal} {\bibinfo
  {journal} {The Journal of Physical Chemistry}\ }\textbf {\bibinfo {volume}
  {81}},\ \bibinfo {pages} {2340} (\bibinfo {year} {1977})}\BibitemShut
  {NoStop}%
\bibitem [{cod()}]{code}%
  \BibitemOpen
  \href {https://github.com/urikm/UtilizeTimeMeasurments.git} {\bibinfo {title}
  {https://github.com/urikm/utilizetimemeasurments.git}}\BibitemShut {NoStop}%
\end{thebibliography}%


\providecommand{\noopsort}[1]{}\providecommand{\singleletter}[1]{#1}%
\begin{thebibliography}{4}%
\makeatletter
\providecommand \@ifxundefined [1]{%
 \@ifx{#1\undefined}
}%
\providecommand \@ifnum [1]{%
 \ifnum #1\expandafter \@firstoftwo
 \else \expandafter \@secondoftwo
 \fi
}%
\providecommand \@ifx [1]{%
 \ifx #1\expandafter \@firstoftwo
 \else \expandafter \@secondoftwo
 \fi
}%
\providecommand \natexlab [1]{#1}%
\providecommand \enquote  [1]{``#1''}%
\providecommand \bibnamefont  [1]{#1}%
\providecommand \bibfnamefont [1]{#1}%
\providecommand \citenamefont [1]{#1}%
\providecommand \href@noop [0]{\@secondoftwo}%
\providecommand \href [0]{\begingroup \@sanitize@url \@href}%
\providecommand \@href[1]{\@@startlink{#1}\@@href}%
\providecommand \@@href[1]{\endgroup#1\@@endlink}%
\providecommand \@sanitize@url [0]{\catcode `\\12\catcode `\$12\catcode
  `\&12\catcode `\#12\catcode `\^12\catcode `\_12\catcode `\%12\relax}%
\providecommand \@@startlink[1]{}%
\providecommand \@@endlink[0]{}%
\providecommand \url  [0]{\begingroup\@sanitize@url \@url }%
\providecommand \@url [1]{\endgroup\@href {#1}{\urlprefix }}%
\providecommand \urlprefix  [0]{URL }%
\providecommand \Eprint [0]{\href }%
\providecommand \doibase [0]{https://doi.org/}%
\providecommand \selectlanguage [0]{\@gobble}%
\providecommand \bibinfo  [0]{\@secondoftwo}%
\providecommand \bibfield  [0]{\@secondoftwo}%
\providecommand \translation [1]{[#1]}%
\providecommand \BibitemOpen [0]{}%
\providecommand \bibitemStop [0]{}%
\providecommand \bibitemNoStop [0]{.\EOS\space}%
\providecommand \EOS [0]{\spacefactor3000\relax}%
\providecommand \BibitemShut  [1]{\csname bibitem#1\endcsname}%
\let\auto@bib@innerbib\@empty
\bibitem [{\citenamefont {Parzen}(1962)}]{KDE}%
  \BibitemOpen
  \bibfield  {author} {\bibinfo {author} {\bibfnamefont {E.}~\bibnamefont
  {Parzen}},\ }\bibfield  {title} {\bibinfo {title} {{On Estimation of a
  Probability Density Function and Mode}},\ }\href
  {https://doi.org/10.1214/aoms/1177704472} {\bibfield  {journal} {\bibinfo
  {journal} {The Annals of Mathematical Statistics}\ }\textbf {\bibinfo
  {volume} {33}},\ \bibinfo {pages} {1065 } (\bibinfo {year}
  {1962})}\BibitemShut {NoStop}%
\bibitem [{\citenamefont {Mart´ınez}\ \emph {et~al.}(2019)\citenamefont
  {Mart´ınez}, \citenamefont {Bisker}, \citenamefont {Horowitz},\ and\
  \citenamefont {Parrondo}}]{GiliNat19}%
  \BibitemOpen
  \bibfield  {author} {\bibinfo {author} {\bibfnamefont {I.~A.}\ \bibnamefont
  {Mart´ınez}}, \bibinfo {author} {\bibfnamefont {G.}~\bibnamefont {Bisker}},
  \bibinfo {author} {\bibfnamefont {J.~M.}\ \bibnamefont {Horowitz}},\ and\
  \bibinfo {author} {\bibfnamefont {J.~M.~R.}\ \bibnamefont {Parrondo}},\
  }\bibfield  {title} {\bibinfo {title} {Inferring broken detailed balance in
  the absence of observable currents.},\ }\href
  {https://doi.org/10.1038/s41467-019-11051-w} {\bibfield  {journal} {\bibinfo
  {journal} {Nat. Commun.}\ }\textbf {\bibinfo {volume} {10}},\ \bibinfo
  {pages} {3542} (\bibinfo {year} {2019})}\BibitemShut {NoStop}%
\bibitem [{\citenamefont {Rold\'an}\ and\ \citenamefont
  {Parrondo}(2012)}]{FR_semiCG_semiAnalytical}%
  \BibitemOpen
  \bibfield  {author} {\bibinfo {author} {\bibfnamefont {E.}~\bibnamefont
  {Rold\'an}}\ and\ \bibinfo {author} {\bibfnamefont {J.~M.~R.}\ \bibnamefont
  {Parrondo}},\ }\bibfield  {title} {\bibinfo {title} {Entropy production and
  kullback-leibler divergence between stationary trajectories of discrete
  systems},\ }\href {https://doi.org/10.1103/PhysRevE.85.031129} {\bibfield
  {journal} {\bibinfo  {journal} {Phys. Rev. E}\ }\textbf {\bibinfo {volume}
  {85}},\ \bibinfo {pages} {031129} (\bibinfo {year} {2012})}\BibitemShut
  {NoStop}%
\bibitem [{\citenamefont {Kim}\ \emph {et~al.}(2020)\citenamefont {Kim},
  \citenamefont {Bae}, \citenamefont {Lee},\ and\ \citenamefont
  {Jeong}}]{NEEP}%
  \BibitemOpen
  \bibfield  {author} {\bibinfo {author} {\bibfnamefont {D.-K.}\ \bibnamefont
  {Kim}}, \bibinfo {author} {\bibfnamefont {Y.}~\bibnamefont {Bae}}, \bibinfo
  {author} {\bibfnamefont {S.}~\bibnamefont {Lee}},\ and\ \bibinfo {author}
  {\bibfnamefont {H.}~\bibnamefont {Jeong}},\ }\bibfield  {title} {\bibinfo
  {title} {Learning entropy production via neural networks},\ }\href
  {https://doi.org/10.1103/PhysRevLett.125.140604} {\bibfield  {journal}
  {\bibinfo  {journal} {Phys. Rev. Lett.}\ }\textbf {\bibinfo {volume} {125}},\
  \bibinfo {pages} {140604} (\bibinfo {year} {2020})}\BibitemShut {NoStop}%
\end{thebibliography}%
\end{document}


\title{Supplemental Material: Utilizing time-series measurements for entropy production estimation in partially observed systems}

\author{Uri Kapustin}
\affiliation{School of Electrical Engineering, Faculty of Engineering, Tel Aviv University, Tel Aviv 6997801, Israel}
\author{Aishani Ghosal}
\affiliation{Department of Biomedical Engineering, Faculty of Engineering, Tel Aviv University, Tel Aviv 6997801, Israel}
\author{Gili Bisker}%
\email{bisker@tauex.tau.ac.il}
\affiliation{Department of Biomedical Engineering, Faculty of Engineering, Tel Aviv University, Tel Aviv 6997801, Israel}
\affiliation{The Center for Physics and Chemistry of Living Systems, Tel Aviv University, Tel Aviv 6997801, Israel}
\affiliation{The Center for Nanoscience and Nanotechnology, Tel Aviv University, Tel Aviv 6997801, Israel}
\affiliation{The Center for Light Matter Interaction, Tel Aviv University, Tel Aviv 6997801, Israel}

\maketitle

\section{Waiting time distributions estimation}

\begin{figure*}[h]
\includegraphics{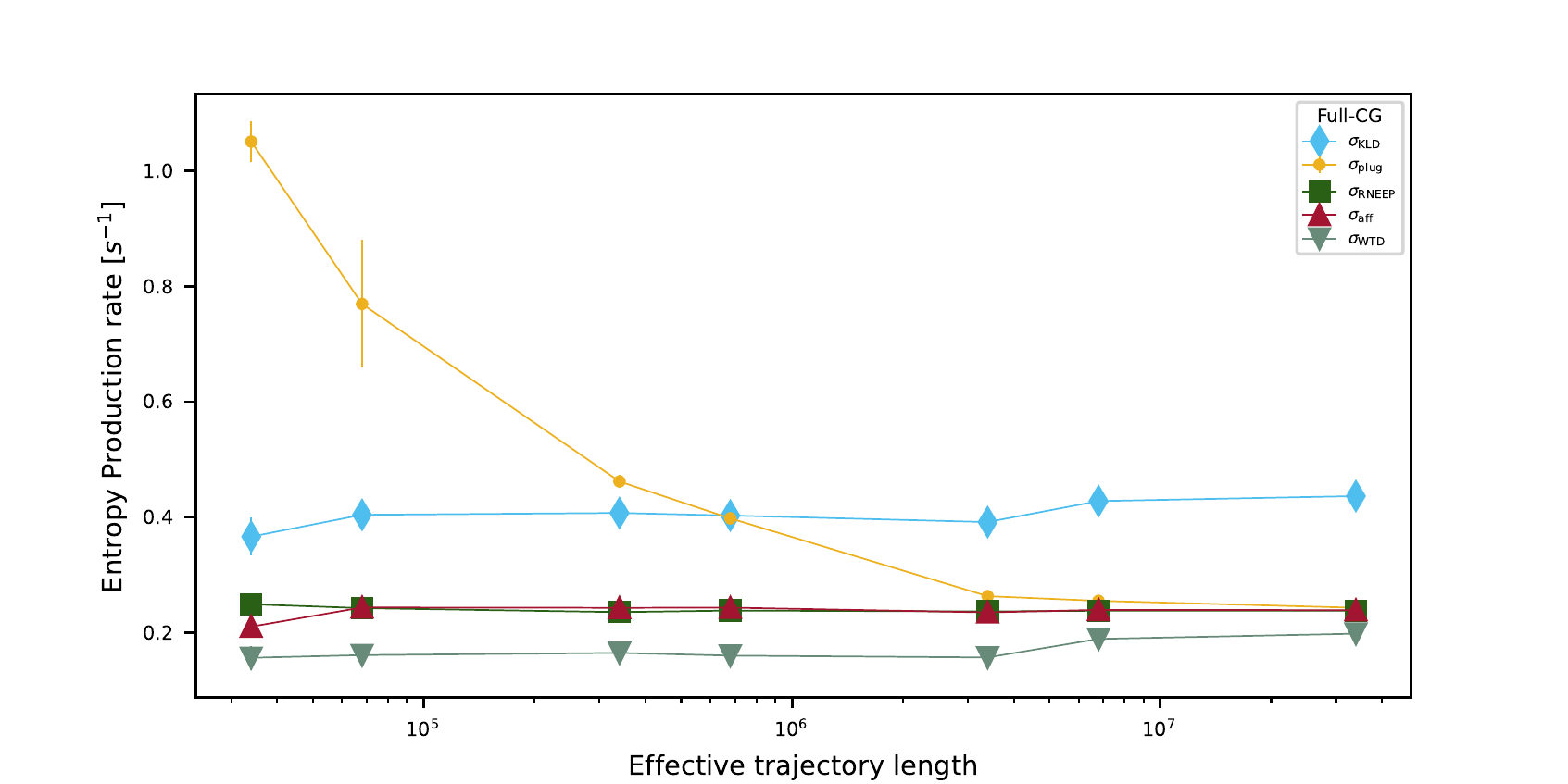}
\caption{\label{fig:lenAnalysis} Convergence of EPR estimators, $\sigma_{\text{KLD}}$ (blue diamonds), $\sigma_{\text{plug}}$ (light orange dots), $\sigma_{\text{RNEEP}}$ (dark green squares), $\sigma_{\text{aff}}$ (red triangles) and $\sigma_{\text{WTD}}$ (gray triangles), as a function of effective trajectory length, 
calculated for the 4-states system and {$x=-0.15$}. The effective trajectory length is calculated after \textit{full-CG}.
}
\end{figure*}

The KLD estimator, $\sigma_{\text{KLD}}$, 
has two contributions, namely,
$\sigma_{\text{aff}}$ and $\sigma_{\text{WTD}}$. 
While $\sigma_{\text{aff}}$ can be directly calculated by counting second-order transitions for second-order semi-Markov processes, $\sigma_{\text{WTD}}$ requires the estimation of continuous functions.
In order to numerically evaluate continuous probability density functions, we use the Kernel Density Estimation (KDE) method {\cite{KDE}}.
In this approach, the estimated function depends on the bandwidth of the kernel, 
and the 
optimal bandwidth is correlated to the sample size. 
In our case, the sample size is the number of observed second-order jumps for each WTD, which can lead to a large variation between sample sizes for different transitions.
We chose three different kernels, each for a different range of sample size, $2\times10^2 - 5\times 10^3$, $5\times10^3 - 10^5$, and $>10^5$.
Sample size of less than $2\times10^2$ was not considered due to the lack of statistics.
The grid size of the KDE was also chosen empirically for optimized convergence to the WTD, considering the computational cost.
The estimation of the WTD is better for longer trajectories, whereas 
short trajectories can result in an inaccurate 
estimation of the EPR due to insufficient statistics (Fig.~\ref{fig:lenAnalysis}). 
Moreover, the required trajectory length for a desired tolerance of the estimation depends on the system parameters \cite{GiliNat19}.


\section{Plug-in Estimator implementation details}
The Plug-in estimator, $\sigma_{\text{plug}}$, is implemented according to \cite{FR_semiCG_semiAnalytical}. Since its calculation includes ratios of probabilities of forward and reverse sequences, a small bias is added to the number of observations of each sequence to avoid probability values being zero. In contrast, for $\sigma_{\text{aff}}$ estimation, unobserved second-order transitions are excluded from the calculation. This inherent difference in the calculations results in the deviation between the $\sigma_{\text{plug}}$ and $\sigma_{\text{aff}}$ values around the stalling force. The value of $\sigma_{\text{plug}}$ as a function of the effective trajectory length can be seen in Fig.~\ref{fig:lenAnalysis}, 
demonstrating the need for sufficient statistics.

\section{RNEEP convergence in 4-state system}
\begin{figure*}[h]
\includegraphics{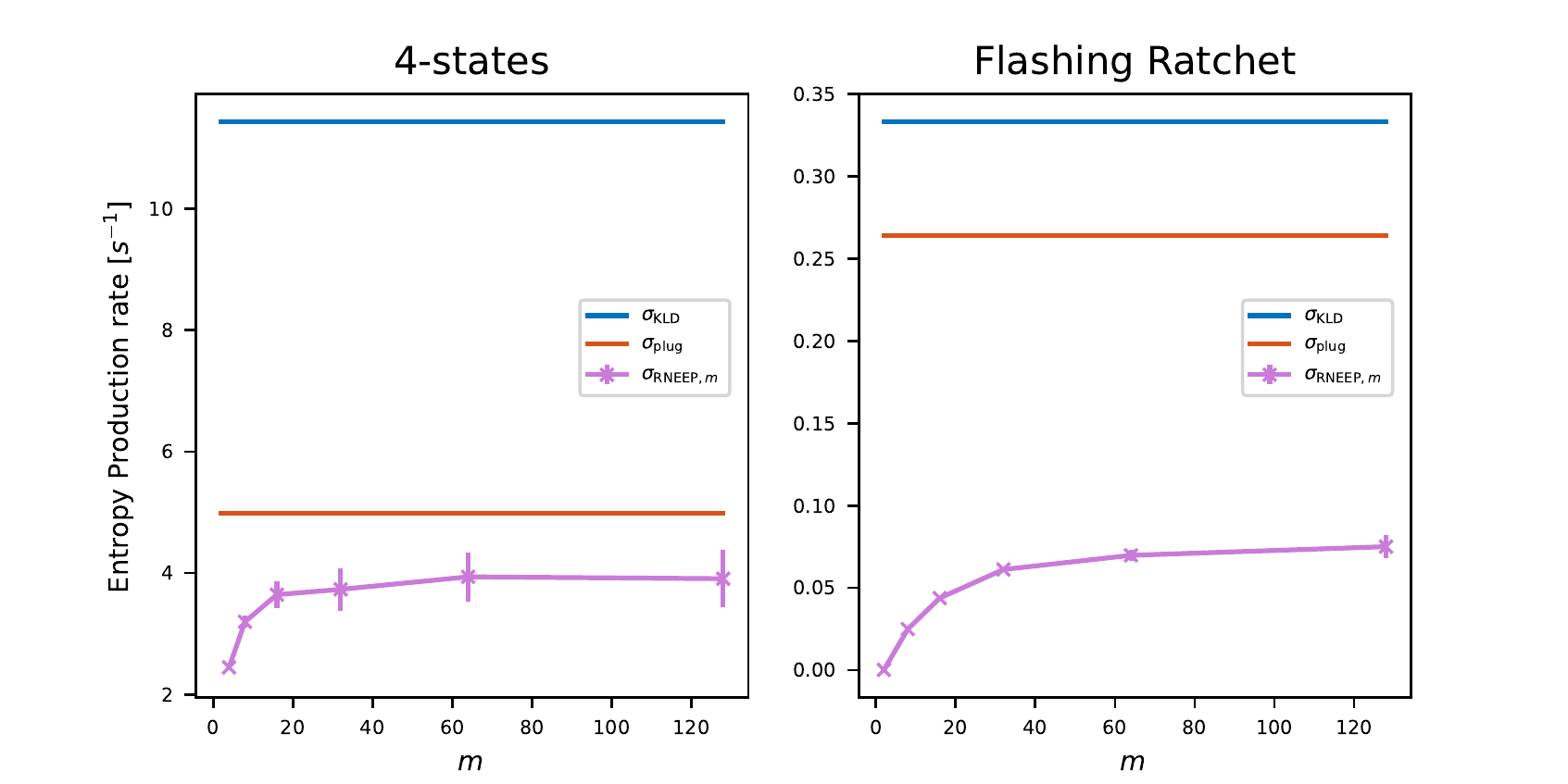}
\caption{\label{fig:converge} Convergence of $\sigma_{\text{RNEEP},m}$ as a function of $m$ (purple) for \textit{semi-CG} data, compared to the values of the
$\sigma_{\text{KLD}}$ (blue diamo), and $\sigma_{\text{plug}}$ (orange), which do not depend of $m$, for the
4-state system
with $x=-2.17$ (left) and for the Discrete Flashing Ratchet
with $V=2$ (right).
}
\end{figure*}

The RNEEP estimator, $\sigma_{\text{RNEEP}}$, is implemented according to \cite{NEEP}, where each data point stems from $12$ training trajectories run in parallel on 8 \textit{Geforce RTX 2080 Ti} GPUs. The $\sigma_{\text{RNEEP},m}$ estimation gives a tighter bound on the total EPR for increasing sequence length, $m$, up to saturation for $m\approx 32$ (Fig.~\ref{fig:converge}) for the $4$-state system.
In contrast,
the flashing ratchet system requires longer sequences for convergence of $m\approx 64$ or longer.


\bibliography{papers}